\documentclass[journal=jacsat,manuscript=article,floatsintext]{achemso}

\usepackage[version=3]{mhchem} 
\usepackage{amsmath}
\usepackage{graphicx}
\usepackage{float}
\usepackage{xcolor}
\usepackage{ulem}

\author{Mashnoon A. Sakib}
\affiliation[UCI]
{Department of Electrical Engineering and Computer Science, University of California, Irvine, CA 92697, USA}
\altaffiliation{Contributed equally to this work}

\author{Naveed Hussain}
\email{naveed.hussain@toyota.com}
\affiliation[UCI]
{Department of Electrical Engineering and Computer Science, University of California, Irvine, CA 92697, USA}
\altaffiliation{Contributed equally to this work}

\author{Mariia Stepanova}
\affiliation[UCI]
{Department of Electrical Engineering and Computer Science, University of California, Irvine, CA 92697, USA}

\author{William Harris}
\affiliation[UCI]
{Department of Physics and Astronomy, University of California, Irvine, CA 92697, USA}

\author{Joshua J. Bocanegra}
\affiliation[UCI]
{Department of Physics and Astronomy, University of California, Irvine, CA 92697, USA}

\author{Ruqian Wu}
\affiliation[UCI]
{Department of Physics and Astronomy, University of California, Irvine, CA 92697, USA}

\author{H. Kumar Wickramasinghe}
\affiliation[UCI]
{Department of Electrical Engineering and Computer Science, University of California, Irvine, CA 92697, USA}
\alsoaffiliation[Second University]
{Department of Materials Science and Engineering, University of California, Irvine, CA 92697, USA}

\author{Maxim~R.~Shcherbakov}
\email{maxim.shcherbakov@uci.edu}
\affiliation[UCI]
{Department of Electrical Engineering and Computer Science, University of California, Irvine, CA 92697, USA}
\alsoaffiliation[Second University]
{Department of Materials Science and Engineering, University of California, Irvine, CA 92697, USA}
\alsoaffiliation[Third University]
{Beckman Laser Institute and Medical Clinic, University of California, Irvine, CA 92612}
\title[An \textsf{achemso} demo]
   {Vacancy-Engineered Phonon Polaritons in a van der Waals Crystal}
  
\begin{document}

\newpage

\begin{abstract}

Phonon-polaritons (PhPs) in low-symmetry van der Waals (vdW) materials confine mid-infrared electromagnetic radiation well below the diffraction limit for nanoscale optics, sensing, and energy control.
However, controlling the PhP dispersion at the nanoscale through intrinsic material properties—without external fields, lithography, or intercalants—remains elusive. Here, we demonstrate vacancy-engineered tuning of PhPs in $\alpha$-phase molybdenum trioxide ($\alpha$-MoO$_3$) via oxygen vacancy formation and lattice strain. Near-field nanoimaging of PhPs in processed  $\alpha$-MoO$_3$
reveals an average polariton wavevector modulation of  $\Delta k/k \approx 0.13 $ within the lower Restrahlen band. Stoichiometric analysis, density functional theory, and finite-difference time-domain simulations show agreement with the experimental results and suggest an induced vacancy concentration of $1\% - 2\%$ along with $(1.2\pm 0.2)\%$ compressive strain, resulting in a non-volatile dielectric permittivity modulation of up to $\Delta \varepsilon / \varepsilon \approx 0.15$. Despite these lattice modifications, the lifetimes of thermomechanically tuned PhPs remain high at $1.2 \pm 0.31$~ps.
These results establish thermomechanical vacancy engineering as a general strategy to reprogram polaritonic response in vdW crystals, offering a new degree of freedom for embedded, non-volatile nanophotonics.


\end{abstract}




\section{Introduction}

Phonon polaritons (PhPs)---quasiparticles arising from the coupling of infrared light with lattice vibrations---offer a compelling route to manipulate electromagnetic fields at the nanoscale.\cite{wu2022manipulating, basov2016polaritons, shiAmplitudePhaseResolvedNanospectral2015, galiffi2024extreme}
In van der Waals (vdW) crystals, PhPs can confine free-space mid-infrared (MIR) light deep below the diffraction limit \cite{wang2024planar,alvarez2020infrared,ma2018plane,hu2023gate,zheng2018highly}, offering applications in 
sub-diffraction imaging\cite{guo2023mid,hu2023gate}, 
long-range hyperlensing\cite{teng2024steering,liang2024manipulation}, 
and engineered photonic states of matter\cite{ma2018plane,chen2020configurable,shen2022hyperbolic}. The key challenge, however, remains in the dynamic control of PhP dispersion through material engineering. While substantial efforts have been made to manipulate PhPs across various engineered and natural material platforms, including by altering the dielectric environment
\cite{chaudhary2019engineering,zhang2021interface,dubrovkin2018ultra} 
incorporating phase change materials\cite{folland2018reconfigurable,fali2019refractive,dai2019phase}, interface-mismatched strain-engineering\cite{zhang2021interface}, and by suspending polar materials in air\cite{yang2022high,shen2022hyperbolic,zheng2022tunable}, a robust method for reconfiguring PhPs in vdW crystals remains elusive.

$\alpha$-phase molybdenum trioxide ($\alpha$-MoO$_3$) emerges as an ideal platform for intrinsically reconfiguring phonon polaritons through lattice engineering.
As a low-symmetry, 2D-layered wide-bandgap ($E_g=2.93$~eV) vdW oxide supporting
low-loss in-plane hyperbolic PhPs\cite{ma2018plane, zheng2019mid, galiffi2024extreme}, $\alpha$-MoO$_3$ exhibits extreme sensitivity to physical and chemical modifications, allowing the exploitation of natural stoichiometry as an alternative dispersion tuning pathway.
Introducing intercalants, such as hydrogen\cite{wang2015reversible}, tin and cobalt\cite{reed2024chemochromism,wu2020chemical}, and isotope enrichment \cite{zhao2022ultralow,zheng2018highly} can substantially modify the PhP dispersion; however, the introduction of foreign atomic species may lead to additional perturbation of the intrinsic homogeneity of the crystal that could adversely impact the PhP 
lifetimes \cite{wu2022manipulating, galiffi2024extreme, taboada2024unveiling, wang2015reversible,reed2024chemochromism,wu2020chemical,zhao2022ultralow}. 
In contrast, 
oxygen vacancies (OVs)---naturally occurring and controllable in $\alpha$-MoO$_3$---offer an attractive avenue to stoichiometric engineering \cite{spevack1992thermal, arnoldy1985temperature}. The stoichiometry of $\alpha$-MoO$_3$ spans from a wide-bandgap oxide with abundant $\ce{Mo}^{6+}$, to intermediate reduced oxides ($\ce{MoO}_{3-x}$, $0 <x< 1$), to semimetallic $\ce{MoO}_{2}$ with a reduced oxidation of $\ce{Mo}^{4+}$\cite{cheng2024recent,de2017molybdenum}. Controlling oxygen content may lead to tunable intrinsic mass composition, allowing targeted modification of optical and acoustic phonons\cite{wu2020chemical} and modulation of the local dielectric permittivity $\varepsilon$, thereby 
offering a route to on-demand control of PhP dispersion.

In this work, we demonstrate that thermomechanically induced OVs and compressive strain enable vacancy-engineered tuning of PhPs in  $\alpha$-MoO$_3$ without lithographic patterning, external fields, or foreign intercalants.
By hot pressing $\alpha$-MoO$_3$ flakes in a pressure- and temperature-controlled environment, we selectively extract the loosely-attached oxygen atoms near the vdW gaps\cite{py1977raman} and induce non-volatile thermal expansion-mismatch-driven strain\cite{hussain2024giant}. Near-field nanoimaging at room temperature with photo-induced force microscopy (PiFM) reveals an average polariton wavevector shift of $\Delta k/k\approx 0.13$ within the lower-Reststrahlen band (L-RB) for processing temperatures between 160--200~$^{\circ}$C
The presence of OVs and strain is confirmed by 
Raman spectroscopy, X-ray photoelectron spectroscopy (XPS), and grazing-incidence X-ray diffraction (GIXRD) measurements. Density functional theory (DFT) and numerical finite-difference time-domain (FDTD) calculations show agreement with the experimental results and suggest oxygen vacancies with concentrations of $1\% - 2 \%$ along with a $(1.2\pm0.2)\%$ compressive strain, yielding 
a nonvolatile dielectric permittivity modulation of up to $\Delta \varepsilon/\varepsilon\approx 0.15$. The lifetimes of tuned phonon-polaritons remain high at 1.2 $\pm$ 0.31 ps with an average lifetime loss of only 29\% compared to pristine material. Our findings establish a previously unexplored tuning mechanism for MIR polaritons, offering a new paradigm for non-volatile reprogramming of light-matter interactions in vdW materials.

\newpage

\begin{figure}[ht!]
\centering
\includegraphics[width=1.\textwidth]{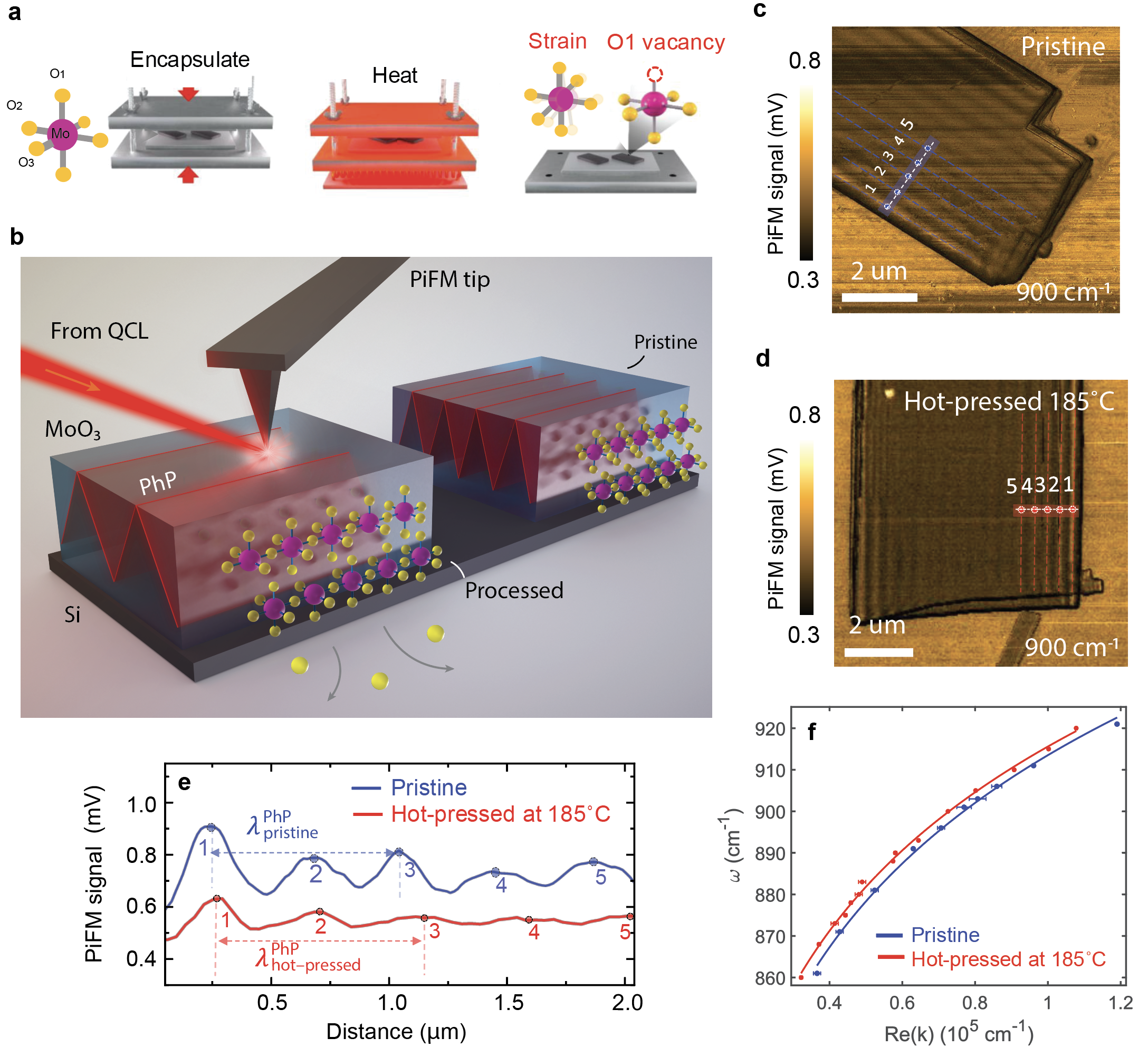}
\caption{\textbf{Thermomechanical hot-press induces strain and oxygen vacancy defects (OVs) in hot-pressed $\ce{\alpha}$-\ce{MoO_{3}} and enables MIR PhP dispersion engineering}. \textbf{a, b}, Schematic illustrations of (\textbf{a}) thermomechanical processing and (\textbf{b}) measurement technique of tip-launched PhPs using the photo-induced force microscopy (PiFM) for $\ce{\alpha}$-\ce{MoO_{3}} flakes on silicon substrates. \textbf{c, d}, PiFM images measured with an excitaion frequency of 900~$\ce{cm^{-1}}$ of (\textbf{c}) a pristine flake ($t_{\ce{pr}}=107 \pm 2$~nm) and (\textbf{d}) a flake hot-pressed at 185~$^{\circ}$C ($t_{\ce{hp}} = 102 \pm 2$~nm). The scale bars are 2 $\mu$m. \textbf{e}, PhP propagation profiles are extracted at 900 $\ce{cm^{-1}}$ shows elongation in PhP wavelength for the hot-pressed flake. \textbf{f}, PhP dispersion $k(\omega)$ measured over the L-RB, $\omega=865 - 915~\ce{cm^{-1}}$. Solid lines are fits to a lower-order polynomial.}
\label{fig1}
\end{figure} 

\vspace{10cm}

\newpage

\section{Results}

\subsection{Nanoimaging PhPs in hot-pressed $\alpha$-MoO$_3$ }

Figure~1a shows a schematic representation of the thermomechanical processing of $\ce{\alpha}$-\ce{MoO_{3}} flakes. Pristine $\ce{\alpha}$-\ce{MoO_{3}} flakes are mechanically exfoliated onto a polished silicon substrate followed by capping with an identical silicon substrate; see Methods. This capped assembly is put inside a pressure device assembly (PDA). The PDA consists of steel bars that generate a localized uniaxial pressure via the top substrate though tightening the screws on the sides of the PDA. The uniaxial pressure exerted by the PDA is found crucial for cutting off oxygen supply during heating the assembly in a muffle furnace up to temperatures ranging from 50 – 400~$^{\circ}$C. Thermomechanical hot-pressing also ensures an interfacial adhesion with the Si substrate during the relative thermal expansion and contraction processes, offering maximum strain transfer\cite{hussain2024giant,hussain2022ultra,hussain2021quantum}. To analyze how the processing induces changes on PhP propagation dynamics and modify their fundamental characteristics, we use PiFM experiments for nanoimaging PhP propagation on pristine and hot-pressed flakes. Fig.~1b illustrates the PiFM technique. Here, a Pt/Ir-coated tip is obliquely illuminated with MIR light to excite tip-launched PhPs. PiFM records the $z$-component of the local field distribution by mechanically detecting the optical force acting on the tip\cite{zheng2019mid,almajhadi2020observation,ambrosio2018selective}. We record the PhP propagation in pristine and hot-pressed $\ce{\alpha}$-\ce{MoO_{3}} flakes within the L-RBs spanning from $865 – 915~\ce{cm^{-1}}$ along the [100] direction. In Fig.~1c,d, we show PiFM images taken at $\omega=900~\ce{cm^{-1}}$ for PhPs in (c) pristine and (d) hot-pressed (185~$^{\circ}$C flakes of similar thicknesses. The scale bars are 2~$\mu$m. Here, the pristine flake, with the dashed blue line in Fig.1c showing a typical cross-section, with a measured thickness $t_{\ce{pr}} = 107 \pm 2$~nm reveals PhP with a wavelength of $\lambda^{\ce{PhP}}_{\ce{pristine}} = 816$~nm and a propagation length of $L_{\ce{pr}} = 1.17 \pm 0.32~\mu$m (see Fig.~S1 and Fig.~S2). In comparison, the hot-pressed flake, with the dashed red line in Fig.1d showing a typical cross-section, with a thickness of $t_{\ce{hp}} = 102 \pm 2$~nm reveals an elongated PhP wavelength with $\lambda^{\ce{PhP}}_{\ce{hot-pressed}} = 892$~nm and a propagation length of $L_{\ce{hp}} = 1.45 \pm 0.38~\mu$m. In Fig.~1e, we show the overlapped PhP profiles for the pristine and 185~$^{\circ}$C hot-pressed flakes recorded from the blue and red stripes in Fig.~1c and d, respectively. Fig.~1e suggests the elongation in PhP wavelength at $\omega= 900~\ce{ cm^{-1}}$. At the same excitation frequency of 900 $\ce{cm^{-1}}$, the 185~$^{\circ}$C hot-pressed flakes, despite having a 5~nm lower thickness, shows an approximate 10\% of increase in PhP wavelength. It is worth noting that the PhP wavevector is typically inversely proportional to the thickness of the flakes, as the decreasing number of layers in thinner flakes are found to inversely affect and increase the oscillator strength that enhances the overall field enhancement through higher PhP wavelength confinement. This effectively shifts the PhP dispersion for thinner flakes to the higher-momentum region\cite{ma2018plane, alvarez2020infrared, wu2022manipulating}. In order to record the dispersion and analyze this PhP propagation characteristic across the L-RBs of $\ce{\alpha}$-\ce{MoO_{3}}, we recorded PhP signals spanning from 865 – 915 $\ce{cm^{-1}}$; see Supplementary Information Figs.~S1 and S2. We found the PhP wavelength for processed flakes is larger across all the measured frequencies compared to those of pristine flakes. We plot the extracted PhP wavevector against the excitation frequencies to map the dispersion characteristics in Fig.~1f for the pristine (blue) and 185~$^{\circ}$C hot-pressed (red) flakes. Fig.~1f suggests that thermomechanically hot-pressing the flakes up to 185~$^{\circ}$C shows an average 7.6\% dispersion shift compared to the pristine flake even though the latter is 5~nm thicker.

\subsection{PiFM characterization of dispersion-modulated PhP propagation}

\begin{figure}[ht!] 
\centering
\includegraphics[width=1\textwidth]{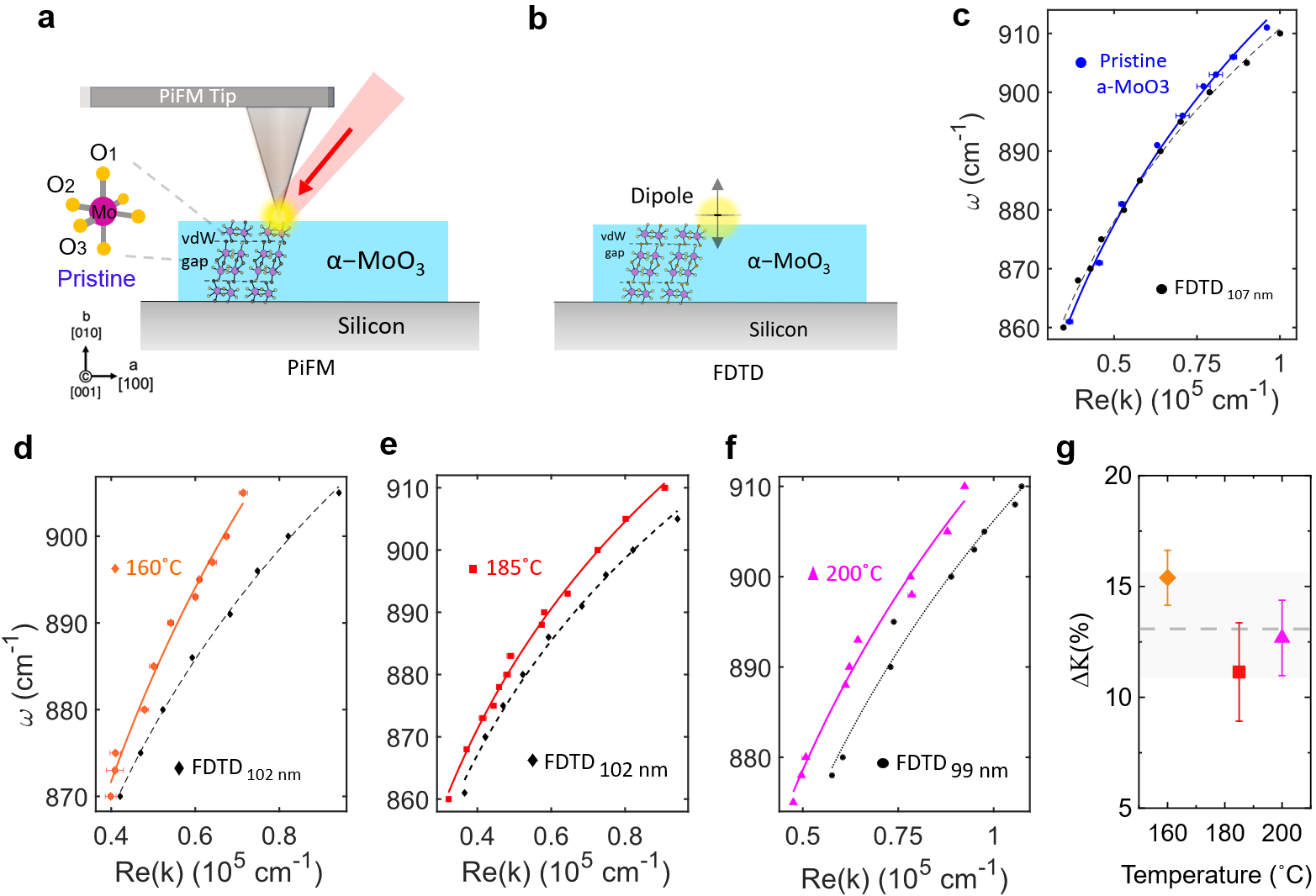}
\caption{\textbf{PiFM characterization of PhPs propagation for pristine and hot-pressed $\ce{\alpha}$-\ce{MoO_{3}} flakes.} \textbf{a,b}, Illustrations of (\textbf{a}) MIR nanoimaging PiFM tip-launched PhPs and (\textbf{b}) finite differece time domain (FDTD) methods with a point-like electric-dipole radiation source launching PhPs in a $\ce{\alpha}$-\ce{MoO_{3}} flake exfoliated on a Si substrate. \textbf{c}, PiFM-recorded dispersion from a pristine 107~nm-thick $\ce{\alpha}$-\ce{MoO_{3}} flake (blue) is plotted against FDTD-simulated dispersion from a flake of the same thickness (black), showing excellent agreement. \textbf{d-f}, PhP dispersion plotted for thermomechanically-proessed $\ce{\alpha}$-\ce{MoO_{3}} flakes with temperatures at (\textbf{d}) 160~$^{\circ}$C, (\textbf{e}) 185~$^{\circ}$C, and (\textbf{f}) 200~$^{\circ}$C with flake thicknesses of $102\pm2$~nm, $102\pm 2$~nm and $99\pm2.5$~nm, respectively. Corresponding FDTD-simulated dispersion from pristine $\ce{\alpha}$-\ce{MoO_{3}} flakes with the same thicknesses are plotted with dashed black curves for comparison. \textbf{g}, Thermomechanical dispersion modulation extracted from the corresponding hot-pressed $\ce{\alpha}$-\ce{MoO_{3}} flakes processed at 160~$^{\circ}$C, 185~$^{\circ}$C and 200~$^{\circ}$C. The error bars represent fitting errors.}
\label{fig2}
\end{figure}

We use PiFM to visualize propagation of PhPs in pristine and hot-pressed $\ce{\alpha}$-\ce{MoO_{3}} flakes processed at 160~$^{\circ}$C, 185~$^{\circ}$C  and 200~$^{\circ}$C; the full datasets for these flakes are givein in Supplementary Information Figs.~S1--3, repectively. Figure~2 summarizes these datasets. Figure~2a,b shows schematic illustrations of (a) mid-IR nanoimaging technique employed by PiFM tip-launched PhPs and (b) Finite differece time domain (FDTD) methodo with a point-like electric-dipole radiation source launched PhPs in pristine $\ce{\alpha}$-\ce{MoO_{3}} membrane exfoliated on a Si substrate; see Methods and Supplementary Information Figs.~S4--5. Figure~2c shows the PiFM-recorded PhP dispersion within the L-RB recorded from a pristine $\ce{\alpha}$-\ce{MoO_{3}} flake with thickness, $t_{\ce{pr}} = 107 \pm 2$~nm (solid blue curve). To build a reference model for quantitative analysis of PhPs, we perform finite difference time domain (FDTD) calculations to simulate the $\ce{\alpha}$-\ce{MoO_{3}} PhPs in L-RBs. The dispersion calculated with FDTD for $t_{\ce{pr}} = 107$~nm is shown in Fig.~2c in black. Fig.~2c suggests an excellent agreement between PiFM and FDTD, with an average error 1.6\% between them mostly stemming from the high-frequency area; we attribute the errors to have originated from the approximative character of the substitution of a real tip with an extended dipole in this approach. This result suggests that the pristine model is a good reference to compare the results obtained with PiFM to the PhP dispersion expected in the pristine flake of the same thickness. This way, any systematic deviations bewteen the PiFM and the FDTD data would point to a tunable response in processed flakes.
 
We perform PiFM experiments for hot-pressed $\ce{\alpha}$-\ce{MoO_{3}} flakes at 160~$^{\circ}$C, 185~$^{\circ}$C and 200~$^{\circ}$C and show their dispersion relationships in Fig.~2d-f. The flake thicknesses for these cases are measured as $t_{\ce{160^{\circ}C}}$ = $t_{\ce{185^{\circ}C}}$ = 102 $\pm$ 2~nm and $t_{\ce{200^{\circ}C}}$= 99 $\pm$ 2.5~nm; for topography data see Supplementaty Information Fig.~S3. To analyze and compare the dispersion relationships, we use FDTD calculations on a pristine $\ce{\alpha}$-\ce{MoO_{3}} flake and plot it in Fig.~2d-e (with $t_{\ce{160^{\circ}C}}$ = $t_{\ce{185^{\circ}C}}$ = 102~nm) and in Fig.~2f (with $t_{\ce{200^{\circ}C}}$ = 99~nm) against the PiFM data. 

It can be seen that hot pressing modulates PhP dispersion towards the low-momentum region by an average 15.4\%, 11.2\% and 12.7\% at 160~$^{\circ}$C, 185~$^{\circ}$C  and 200~$^{\circ}$C, respectively, with an average 13.1\% dispersion modulation, as shown in Fig.~2g. This ability to modify the PhP wavelength presents the potential for achieving on-demand dispersion configurability in vdW materials\cite{ma2018plane, chen2020configurable, shen2022hyperbolic}. The highest wavelength elongation for L-RBs in hot-pressed $\alpha$-MoO3 is observed to be up to 24\%. We attribute this elongation to OV and strain-induced non-volatile increase in the dielectric permittivity for hot-pressed $\ce{\alpha}$-\ce{MoO_{3}}. To characterize the PhP propagation dynamics, we extract the PhP lifetimes, $\tau$, by fitting the PhP linescans to an exponentially decaying sinusoidal function; see Supplementary Information Figs.~S1-3. The calculated PhP lifetimes in thermomechanically processed $\ce{\alpha}$-\ce{MoO_{3}} flakes are $1\pm0.2$~ps, $1.4\pm0.2$~ps, and $1.1\pm0.3$~ps  for 160~$^{\circ}$C, 185~$^{\circ}$C  and 200~$^{\circ}$C, respectively. These PhP lifetimes are comparable to those in pristine $\ce{\alpha}$-\ce{MoO_{3}} flakes, measured to be $1.2 \pm 0.31$~ps, hinting at resilience of PhP propagation after the introduction of OVs and strain into the $\ce{\alpha}$-\ce{MoO_{3}} crystal homogeneously across the L-RB band. For lifetimes measured at L-RB frequencies spanning from $871-915~\ce{cm^{-1}}$, the PhP propagation characterization suggests a reasonable average 29\% loss, establishing hot-pressing as a robust method to tune the PhP propagation when compared to other reported works that involve modification of the homogeneity of the crystal \cite{zheng2018highly,wu2020chemical, zhao2022ultralow, reed2024chemochromism}.

\subsection{Stoichiometry characterization of hot-pressed $\alpha$-MoO3}

To establish the nature of the observed PhP tunability, we extensively characterize pristine and hot-pressed flakes using Raman spectroscopy, X-ray photoelectron spectroscopy (XPS), grazing-incidence X-ray diffraction (GIXRD), and energy-dispersive X-ray spectroscopy (EDS), allowing us to establish OVs and strain as the primary reasons for dielectric permittivity modulation.

Fig.~3a shows the Raman spectra of pristine and hot-pressed flakes at extended temperatures ranging from 50 to 400~$^{\circ}$C to evaluate thermomechanical effects on its phononic structure. The spectrum suggests vibrational bands that correspond to the stretching modes between 1000 and 600 $\ce{cm^{-1}}$, deformation modes between 600 and 400 $\ce{cm^{-1}}$, and lattice modes below 200 $\ce{cm^{-1}}$. All samples displayed distinct and robust bands at 117, 129, 245, 284, 291, 338, 338, 665, 819, and 996 $\ce{cm^{-1}}$. A typical $\ce{\alpha}$-\ce{MoO_{3}} crystal comprises of a double layer of linked and deformed $\ce{MoO_{6}}$ octahedra and is thermodynamically stable\cite{sun2023lattice,amba2024renewable}. The relatively weak peak at 996~$\ce{cm^{-1}}$ is the A$_{\ce{g}}$ mode attributed to the asymmetric $\ce{Mo^{6+}= O(1)}$ stretching mode of terminal oxygen along the b-axis. The band at 819 $\ce{cm^{-1}}$ is the most intense band that corresponds to the symmetric stretching mode of doubly coordinated oxygen $\ce{Mo^{2}= O(3)}$, which originates from the oxygen shared by the two $\ce{MoO_{6}}$ octahedra and is sensitive to oxygen vacancies and defects\cite{illyaskutty2014alteration}. The full width at half maximum (FWHM) of this peak provides crucial information regarding the presence of OVs in $\ce{\alpha}$-\ce{MoO_{3}} flakes. The peak broadens as hot-pressing temperature increases, with the FWHM increasing from 8.2 $\ce{cm^{-1}}$ (for pristine) to 15.12 $\ce{cm^{-1}}$ (for flakes h.p. at 350~$^{\circ}$C); see Supplementary Information Fig.~S8. This inhomogeneous peak broadening could result from a decreased phonon lifetime, which is typically driven by an increase in the concentration of OVs, i.e., a significant decrease in the oxygen-to-metal ratio. Probing the $\ce{T_{b}}$ ($\ce{A_{g}}$/$\ce{B_{1g}}$), the translational chain mode at 158.4 $\ce{cm^{-1}}$ determines the lattice strain, while the $\ce{T_{c}}$ $\ce{(B_{2g}/B_{3g})}$ mode comprising of 284 $\ce{cm^{-1}}$ $\ce{(B_{2g})}$ and its shoulder at 291 $\ce{cm^{-1}}$ $\ce{(B_{3g})}$  assesses the presence of OVs. The results suggest that hot pressing of the pristine $\ce{\alpha}$-\ce{MoO_{3}} flakes affects its phonon modes related to the lattice and valence state, introducing temperature-induced OVs and eventually resulting in lattice strain in $\ce{\alpha}$-\ce{MoO_{3}}.  

\begin{figure} [t!]
\centering
\includegraphics[width=1\textwidth]{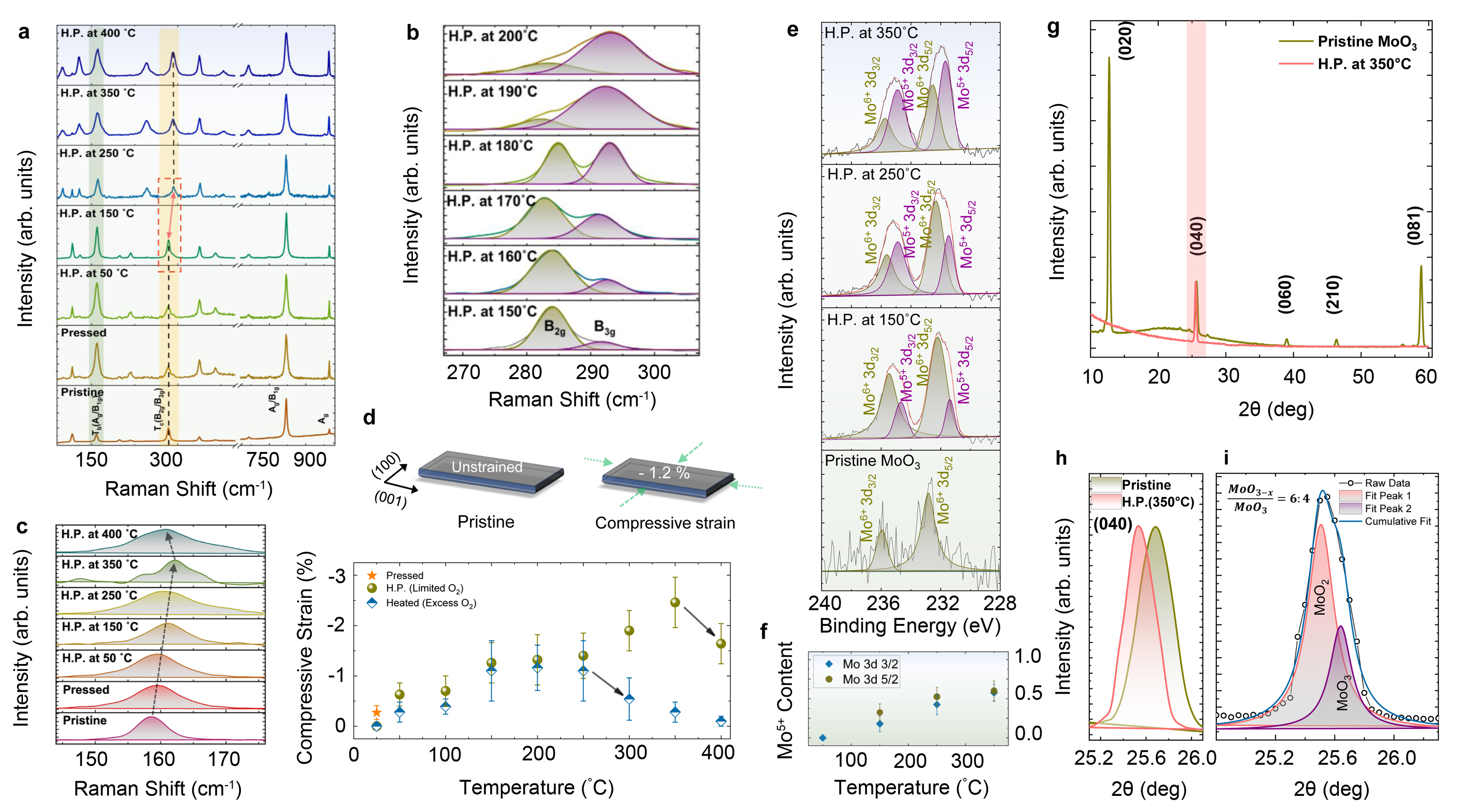}
\caption{\textbf{Stoichiometry characterization of hot-pressed $\ce{\alpha}$-\ce{MoO_{3}}}. \textbf{a}, Raman spectra of the pristine, pressed and h.p. $\ce{\alpha}$-\ce{MoO_{3}} at 50, 150, 250, 350, and 400~$^{\circ}$C, with $\ce{T_{b}}$ and $\ce{T_{c}}$ mode highlighted around ~150~$\ce{cm^{-1}}$ and ~290~$\ce{cm^{-1}}$ with shaded regions, respectively.  \textbf{b},  Magnified Raman spectra (normalized) of $\ce{T_{c}}$ $\ce{(B_{2g}/B_{3g})}$ mode acquired by h.p. $\ce{MoO_{3}}$ from 150 to 200~$^{\circ}$C shows a band transition with a decrease in oxygen-to-metal ratio, highlighted by orange arrow in (a).  \textbf{c, d},  Blue shift in $\ce{T_{b}}$ $\ce{(A_{g}/B_{1g})}$ (translational chain) mode with h.p. temperature, suggesting lattice strain engineering. Comparison of compressive strain induction in only heated (without encapsulation) and thermomechanically-processed h.p. flakes with variation in temperatures. The gray dashed region in (c) hints a $(1.2\pm 0.2)\%$ compressive strain originating from the thermomechanical processing temperatures measured at 150~$^{\circ}$C and 200~$^{\circ}$C.  \textbf{e}, Mo$_{3d}$ scan of pristine, and h.p. flakes at 50, 150, 250, 350~$^{\circ}$C, obtained from XPS to probe the processed $\ce{\alpha}$-\ce{MoO_{3}} stoichiometry.  \textbf{f},  $\ce{Mo^{5+}}$ content in $\ce{\alpha}$-\ce{MoO_{3}}, calculated using peak area fitting of Mo $3d_{3/2}$ and $3d_{5/2}$ peaks.  \textbf{g}, GIXRD pattern of the pristine, and h.p.~$\ce{MoO_{3}}$ at 350~$^{\circ}$C.  \textbf{h},  Comparison of the (040) diffraction peak (normalized) of pristine and h.p. \ce{MoO_{3}} at 350~$^{\circ}$C, showing a shift by 0.4.  \textbf{i}, Deconvolution of (040) diffraction peak of h.p. $\ce{\alpha}$-\ce{MoO_{3}} yields a ratio of $\ce{MoO_{2}}$ and \ce{MoO_{3}} of 6:4.}
\label{fig3}
\end{figure}

In Fig.~3b,c, we estimate the lattice strain for h.p. flakes. Fig.~3b suggests a consistent blue shift at 158.4 $\ce{cm^{-1}}$ for samples hot-pressed up to 350~$^{\circ}$C before relaxing back to its position at 400~$^{\circ}$C.
The strain and its type are calculated as 
$
\delta (\%) = (\omega_{\mathrm{T_b}}^{\mathrm{pristine}} - \omega_{\mathrm{T_b}}^{\mathrm{h.p.}})/ \omega_{\mathrm{T_b}}^{\mathrm{pristine}} \times100,
$
 where $\omega_{\mathrm{Tb}}^\mathrm{pristine}$ and $\omega_{\mathrm{Tb}}^{\mathrm{h.p.}}$  are the frequencies of the $\ce{T_b}$ phonon mode corresponding to pristine and hot-pressed $\alpha$-$\ce{MoO}_{3}$, respectively. We attribute this $\ce{T_b}$ shift to the compressive strain caused by the coefficient of thermal expansion (CTE) mismatch at the interface between silicon substrate and $\alpha$-$\ce{MoO}_{3}$ flakes, while the relaxation is caused by the slippage that supersedes the interfacial adhesion above 350~$^{\circ}$C. 
We anticipate that as the substrate transfers compressive strain, it may increase the interlayer adhesion between the silicon substrate and $\ce{\alpha}$-\ce{MoO_{3}} flakes. This is due to the encapsulated 'sandwich' assembly of $\ce{\alpha}$-\ce{MoO_{3}} flakes by the top and bottom silicon substrate. The uniaxial pressure provided by the top substrate is critical. To study the impact of uniaxial pressure on compressive strain, we characterized heated samples without any uniaxial pressure from the upper substrate. Fig.~3c shows the estimated compressive strain values for hot-pressed (green) and bare-heated samples (navy). In bare-heated samples, the lack of uniaxial pressure results in interfacial slippage at approximately 250~$^{\circ}$C, which restricts the maximum strain transfer to $-(1.2\pm 0.2)\%$ due to the relative CTE mismatch. In contrast, interfacial slippage in hot-pressed samples was observed at elevated temperatures (350~$^{\circ}$C), resulting in a two-fold increase in maximum strain transfer –$(2.4\pm 0.2)\%$. We anticipate that the strain observed in the lattice may be due not only to the CTE mismatch but also to the introduction of OVs.         

Additionally, the oxygen vacancies in $\ce{\alpha}$-\ce{MoO_{3}} can be detected by variation in the ratio of Raman mode intensities of $\ce{B_{2g}}$ at 284 $\ce{cm^{-1}}$ and $\ce{B_{3g}}$ at 291 $\ce{cm^{-1}}$ as $\ce{B_{2g}/B_{3g}}$ \cite{liu2018huge,de2020raman}. In pristine $\ce{\alpha}$-\ce{MoO_{3}}, the weak intensity of the $\ce{B_{3g}}$ shoulder peak corresponds to an intrinsically small amount of oxygen vacancies. We observed a consistent increase in this peak's intensity, making it the primary marker of OV creation. Furthermore, we performed XPS to identify the stoichiometry of pristine and h.p. $\ce{\alpha}$-\ce{MoO_{3}}. We show the Mo 3d scans of pristine, h.p. at 150, 250, and 350~$^{\circ}$C samples in Fig.~3e. For pristine $\ce{\alpha}$-\ce{MoO_{3}}, Mo $\ce{3d_{3/2}}$ and Mo $\ce{3d_{5/2}}$ are located at binding energies of 235.92 and 232.82~eV, respectively, which corresponds to pure $\ce{MoO_{3}}$ stoichiometry with $\ce{Mo^{6+}}$ oxidation state with hexavalent formal molybdenum ions\cite{alsaif2014tunable,swiatowska2008li}. As the temperature rose to 350~$^{\circ}$C, the h.p. flakes demonstrated a gradual rise in the proportion of the $\ce{Mo^{5+}}$ state (purple peaks),indicating a shift in the composition towards lower oxidation and $\ce{MoO_{3-x}}$. The calculations on fitted peak areas reveal that the h.p. flakes represent a hybrid system, with the $\ce{Mo^{5+}}$ content in $\ce{MoO_{3-x}}$ increasing from 0\% at 25~$^{\circ}$C to approximately 60\% for the flakes hot-pressed at 350~$^{\circ}$C, as shown in Fig.~3f. XPS results complement the Raman results and suggest that the chemical stoichiometry can be controlled by the introduction of OVs with the increase in h.p.~temperature. 
Additionally, GIXRD results shown Fig.~3g-i and EDS results summarized in Supplementary Information Figs.~S10-11 
quantify the transition from pristine $\ce{\alpha}$-\ce{MoO_{3}} to reduced $\ce{\alpha}$-\ce{MoO_{3-x}} in hot pressed flakes.  
The results in Fig.~3g-i indicate the transition from polycrystalline to monocrystalline phase in h.p. flakes as there were no peaks other than (040) along the basal plane at 25.53~$^{\circ}$, compared to its pristine counterpart. However, Fig.~3i shows a shift of 0.4~$^{\circ}$ in (040) peak, which indicates the existence of $\ce{MoO_{3-x}}$\cite{kim2017oxygen,zhang2021high}. Gaussian peak fitting suggested that the $\ce{MoO_{2}}$: $\ce{MoO_{3}}$ ratio in a 350~$^{\circ}$C h.p. flake was 6:4 (Fig.~3i). XPS results indicate a partial reduction of $\ce{\alpha}$-\ce{MoO_{3}} ($\ce{Mo^{6+}}$) to $\ce{\alpha}$-\ce{MoO_{3-x}} ($\ce{Mo^{5+}}$) as well. 

To summarize, the stoichiometric analysis of the hot-pressed $\ce{\alpha}$-\ce{MoO_{3}} confirms the controllable creation of OVs and compressive strain in a wide range of processing temperatures, highlighting their potential roles in the observed PhP modulation.

\subsection{DFT calculations of strain- and OV-induced index modulation}
\begin{figure} [ht!]
\centering
\includegraphics[width=.8\textwidth]{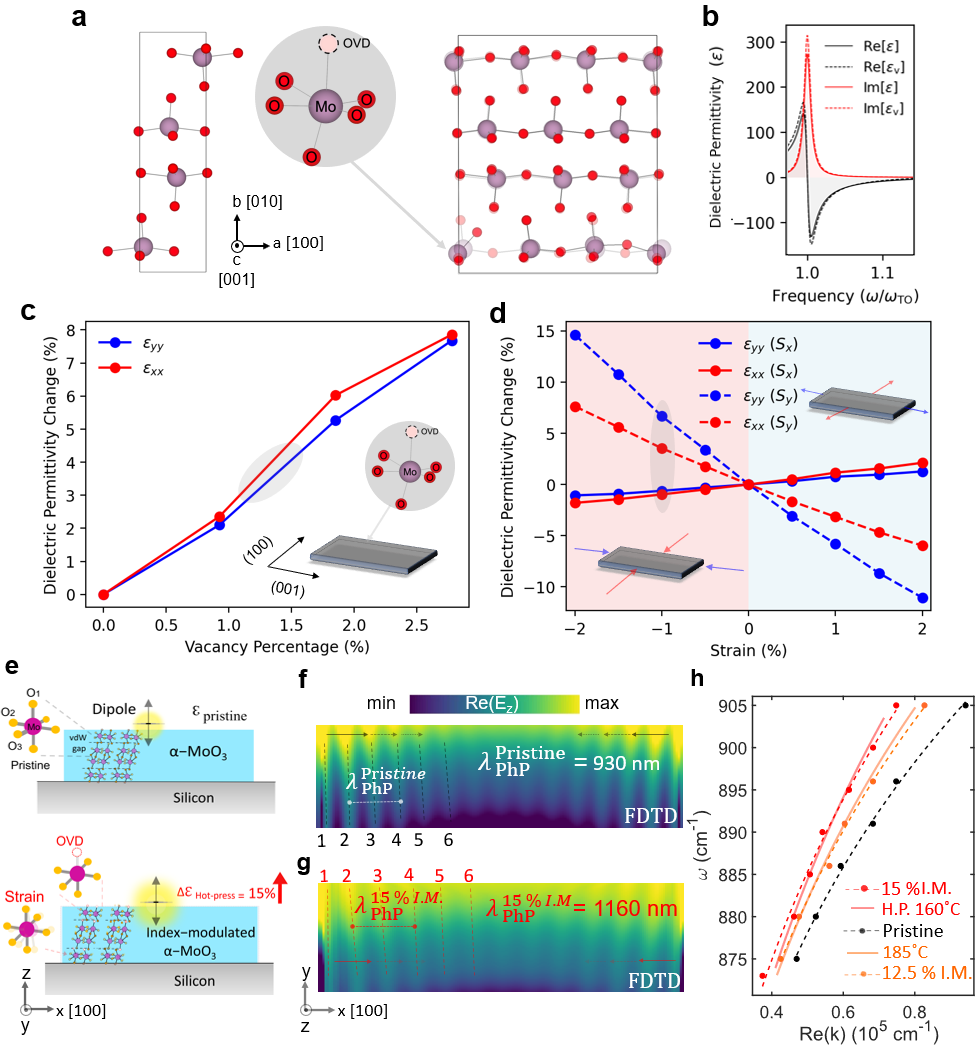}
\caption{ DFT and FDTD calculations of dielectric permittivity modulation in hot-pressed $\ce{\alpha}$-\ce{MoO_{3}}. (a) Crystal structure of relaxed $\ce{\alpha}$-\ce{MoO_{3}} with an oxygen vacancy at the O1 site near vdW gap. (b) The corresponding IR dielectric response in the [100] direction, where solid lines represent the dielectric permittivity of pristine $\ce{\alpha}$-\ce{MoO_{3}}, and dashed lines correspond to that of the oxygen deficient $\ce{\alpha}$-\ce{MoO_{3}}. Dielectric functions are plotted with respect to ($\omega$/$\omega_{\mathrm{TO}}$), where $\omega_{\mathrm{TO}}$ is the transverse optical phonon frequency of $\ce{\alpha}$-\ce{MoO_{3}}. (c-d) Static dielectric constants analyzed as functions of (c) OV concentration, where vacancies are positioned near the vdW gap, and (d) applied strain along the $x$ [100] and $y$ [001] directions, including both compressive (-) and tensile (+) strain. DFT calculations of OV-induced $\ce{\alpha}$-\ce{MoO_{3}} suggest dielectric changes of up to 8\% with OV concentrations reaching up to 3\%. (e) Illustration of pristine and hot-pressed permittivity modeling using FDTD.(f-g) Numerically simulated field distributions (real part of the out-of-plane component of the electric field, $\ce{Re(E_{z}})$ along the X-Y plane for the pristine (in Fig.~4f) and for the index-modulated $\ce{\alpha}$-\ce{MoO_{3}} (in Fig.~4g). (h) Numerically calculated dispersion of PhPs for a 102~nm $\ce{\alpha}$-\ce{MoO_{3}} membrane for pristine and for permittivity modulation of 12.5~\% and 15~\%.}

\label{fig3}
\end{figure}

To explain the dispersion-modulated PhP wavelength elongation, we consider both thermomechanically induced strain and OVs as the primary causes, as evidenced by the stoichiometric analysis. We employ density functional theory (DFT) and FDTD methods to analyze and compare the synergistic effects of strain- and OV-induced index modulation across various OV concentrations and crystal strain levels. The results suggest that the elongation of $\lambda^{\mathrm{PhP}}$ can be attributed to an increase in the static dielectric constants of hot-pressed $\ce{\alpha}$-\ce{MoO_{3}}, driven by both OVs and lattice strain. The static dielectric shifts are assumed to combine linearly under small material modulations, such that  $\Delta \varepsilon = \Delta \varepsilon_{\mathrm{OV}} + \Delta \varepsilon_{\mathrm{Strain}}$. 

DFT calculations estimate the static dielectric constants of $\ce{\alpha}$-\ce{MoO_{3}} with various defect concentrations utilized a $3\times3\times1$ supercell, with oxygen vacancies positioned near the vdW gap as shown in Fig.~4a, as these sites are the most energetically shallow. For the details of DFT calculations, refer to Methods. Up to three defects per supercell were considered. While the relative spatial position of the defects does not change the direction of the dielectric trend, it does affect the magnitude of the shifts; see Supplementary Information Fig.~S12. This site-specific variation is expected to average out under typical experimental conditions. Moreover, as shown in Fig.~4c, the results indicate an increase in the dielectric response with an approximately linear trend. A dielectric permittivity modulation of up to $\sim 8$\% is observed for up to $\sim 3$\% oxygen vacancy concentration. We also found strain to influence the dielectric properties, particularly along the [100] direction. This dependence is attributed to the strong modulation of the conduction band near the gamma point from [100] strain, which, in turn, modulates the band gap; see Supplementary Information Fig.~S13. In flakes subjected to hot-pressing treatments at 160~$^{\circ}$C, 185~$^{\circ}$C, and 200~$^{\circ}$C, dispersion shifts of up to 15.4\% were measured. These shifts could correspond to an estimated oxygen vacancy concentration ranging between 1 - 2\% of oxygen atoms, assuming linearity beyond the calculated range. Alternatively, compressive strain along the [100] direction of approximately -$(1.2\pm 0.2)\%$ may explain the experimental shifts observed. 

We use FDTD to simulate the effects of increased permittivity on $\lambda_{PhP}$ in hot-pressed $\ce{\alpha}$-\ce{MoO_{3}}. Here, we model the OV and strain-induced index modulation by modifying $\ce{\alpha}$-\ce{MoO_{3}}'s intrinsic static dielectric constants. We show this illustration of pristine and OV and strain-induced permittivity modeling using FDTD in a schematic using Fig.~4e. We run FDTD with a 102~nm-thick $\ce{\alpha}$-\ce{MoO_{3}} flakes with both the pristine and the 15~\% index-modulated $\ce{\alpha}$-\ce{MoO_{3}} model; see Supporting Information Fig.~S5. We show the simulated real part of the out-of-plane component of the electric field Re$(E_{z})$ along the X-Z plane for the pristine and for OV-populated and strained $\ce{\alpha}$-\ce{MoO_{3}} Fig.~4f and g, respectively. Simulated Re$(E_{z})$ shown in Fig.~4f-g represents PhPs propagation at $\omega$ = 890 $\ce{cm^{-1}}$. At 890 $cm^{-1}$, FDTD simulations suggest that the PhP propagation along the [100] direction reveal a 24~\% increase in $\lambda_{\mathrm{PhP}}$ elongation for the 15~\% index-modulated $\alpha$-MoO3 membrane, compared to the pristine $\ce{\alpha}$-\ce{MoO_{3}} flake. This wavelength elongation, in turn, shifts the dispersion towards the lower-momentum region. We run FDTD calculations for other frequencies ranging from 870 $\ce{cm^{-1}}$ - 905 $\ce{cm^{-1}}$ and calculate the dispersion for the cases of pristine and thermomechanically processed $\ce{\alpha}$-\ce{MoO_{3}} flakes. In Fig.~4h, we show these simulated dispersion curves for the 102~nm $\ce{\alpha}$-\ce{MoO_{3}} pristine flake (dashed black curve) and for a permittivity modulation of 12.5~\% (dashed orange curve) and 15~\% (dashed red curve), respectively. The orange and red solid curves represent the dispersion recorded from PiFM nanoimaging for hot-pressed samples at 160~$^{\circ}$C and 185~$^{\circ}$C, respectively. We found excellent agreement between the dispersion curves plotted for the cases of permittivity-modulated FDTD calculations and PiFM-recorded dispersions plotted from OV and strain-induced hot-pressed flakes. This suggests that for a 102~nm pristine $\ce{\alpha}$-\ce{MoO_{3}}, a 12.5\% and a 15~\% increase in the static dielectric components along the two $x$ and $y$ directions can model the response of hot-pressed flakes processed at 185~$^{\circ}$C and 160~$^{\circ}$C, respectively. 




\section{Discussion}

To compare the modulation capability of thermomechanical engineering of $\ce{\alpha}$-\ce{MoO_{3}} in terms of controlling phonon-polaritons, we have compared each approach in Supplementary Information Fig.~S4. We have demonstrated an effective stoichiometric engineering approach to permanently manipulate PhPs in $\ce{\alpha}$-\ce{MoO_{3}} by the oxygen vacancy and strain-induced perturbations of the lattice resulting in a highly tunable non-volatile modulation of dielectric permittivity. Specifically, in terms of dispersion tunability, our approach demonstrates the highest tunability across the reported stoichiometric engineering approaches. Moreover, the figures of merits related to the PhP propagation, such as quality factor and damping rate, remain quantitatively in good standing among all the reported PhP engineering methods in $\ce{\alpha}$-\ce{MoO_{3}}.

Additional efforts are required to extend the capability and realize the full potential for controlled PhP manipulation. Ramping up the thermomechanical processing temperature for $\ce{\alpha}$-\ce{MoO_{3}} toward high-temperature regimes, such as 400 - 700~$^{\circ}$C and characterizing its effects on the upper-Reststrahlen bands may help further insight towards a better understanding the effects of oxygen vacancies and strain onto the PhPs. Expanding on these insights could help the community sketch a more generalized mechanism for developing controlled oxygen-vacancy induction methods in polar crystals. This could potentially lead to further research that may specifically relate to not only the class of oxygen-rich polar crystals such as $\ce{\alpha}$-\ce{V_{2}O_{5}}\cite{taboada2020broad}, $\ce{\beta}$-\ce{GaO}\cite{matson2023controlling},\ce{LiV_{2}O_{5}}\cite{f2024observation},\ce{SrTiO_{3}}\cite{xu2024highly} but also their heterostructures\cite{zhou2023thermal}. 
Our proof-of-concept may help to develop a stoichiometric approach that provides a fundamental route toward controlling the constituents \textit{in situ} at the interface between two or more general classes of oxygen-rich systems. The heterostructures may also be controlled externally though voltage and direct injection of carriers with electrical gating, 2D heterostructures such as graphene-gated\cite{hu2022doping,ruta2022surface,zhou2023gate} devices for doping-driven topological polaritons may provide additional tuning knobs to further explore and extend the tuning capabilities for advancing the development of modulation schemes for doping-driven topological polaritons in actively tunable devices. This exciting research avenue may provide a rich playground for exploring the topological transitions of PhPs in these much sought-after classes of polar materials and leveraging on the new polariton physics. 


Here, we propose a thermomechanical approach to highly tunable phonon-polaritons in $\ce{\alpha}$-\ce{MoO_{3}} via controlled oxygen vacancy formation and compressive strain, which achieve permanent changes in the dielectric function without lithography, external fields, or intercalants. 
Photo-induced microscopy reveals an
average polariton wavevector shift of $\Delta k/k=0.13$ 
within the lower-Restrahlen band for processing temperatures between 160--200~$^{\circ}$C. Stoichiometric analysis 
confirms oxygen vacancies with concentrations ranging from $1\% - 2\%$ along with a compressive strain $(1.2\pm 0.2)\%$, in agreement with density functional theory and finite-difference time-domain simulations, yielding a non-volatile modulation of the dielectric permittivity of up to 15\%. Despite the structural changes, the lifetimes of vacancy-engineered phonon-polaritons 
are measured at 1.2 $\pm$ 0.31 ps with an average loss in lifetime on only  29\% compared to the pristine. Our findings
establish thermomechanical vacancy engineering as a viable route for reprogramming MIR light--matter interactions in vdW materials, opening new possibilities for integrated, non-volatile polaritonic nanophotonics.

\newpage
\section{Methods}

\subsection{Thermomechanical processing of $\ce{\alpha}$-\ce{MoO_{3}} flakes}
We mechanically exfoliated high-quality $\ce{\alpha}$-\ce{MoO_{3}} from bulk $\ce{\alpha}$-\ce{MoO_{3}} crystals (2D Semiconductor Inc.) directly onto precleaned and flat (10×10 \ce{mm^{2}}) Si substrates. Following the exfoliation, we use another identical Si substrate to sandwich the $\ce{\alpha}$-\ce{MoO_{3}}. The assembly was then placed in a home-built pressure device to apply mild uniaxial pressure onto the flakes; see Supplementary Information, Fig.~ S1b. We then placed the overall pressure device assembly into a commercial benchtop muffle furnace (Thermo Scientific) and heated it for 30 minutes at various temperatures in an ambient atmosphere.

\subsection{PiFM nanoimaging measurements}
A Molecular Vista Inc. VistaScope microscope was connected to a Block Engineering LaserTune quantum cascade laser (QCL) system, with wave number resolution of 0.5 $cm^{-1}$ and a tunable range from 782 to 1920 $cm^{-1}$. During the operation, the microscope utilized NCH - PtIr 300 kHz cantilevers from Molecular Vista and operated in sideband mode at a QCL intensity of 5\%.

\subsection{FDTD simulations}
We performed full-wave electromagnetic simulations using a finite-difference time-domain (FDTD) method as implemented in Ansys Lumerical FDTD software. The boundary conditions along the $x$-, $y$- and $z$-directions were set with perfectly matched layers. To excite and launch highly confined PhPs, we use a point-like electric dipole source polarized along the $z$-direction. The dipole was positioned at a height of 100-150~nm from the uppermost surface of the target $\ce{\alpha}$-\ce{MoO_{3}}. We record the $\ce{Re(E_{z}})$ at a distance of 10~nm on top of the uppermost surface. In this method, we scan the dipole across the topmost surface of the flake and record Re($E_{z}(x,y)$).We extract the dispersion contours using a fast Fourier transform of the recorded Re($E_{z}$). Here, it is noted that in PiFM, the QCL uses a p-polarized light for illuminating the metal tip kept 30-50 nm away from the top surface of the target flakes. As a result, the collected polaritonic electric fields in the form of photo-induced force (PiF signal) are predominantly concentrated along the $z$-direction. Such a dipole excitation mechanism allows us to mimic the PiFM excitation and collection scheme. The permittivity of the $\ce{\alpha}$-\ce{MoO_{3}} flakes has been modeled with a Drude-Lorentz model as reported in previous reports\cite{alvarez2020infrared}. For the details of the permittivity modeling in the cases of pristine and hot-pressed $\ce{\alpha}$-\ce{MoO_{3}} flakes, see Supplementary Information Figs.~S6-S8.

\subsection{Raman microscopy}
A confocal Raman microscope (Renishaw Inc.), equipped with an objective lens (Nikon Plan Fluor 50 X, NA=0.4) and a 532~nm unpolarized laser source (22 mW, 50× objective, spot size 1 $\mu$m) in ambient conditions was used to acquire micro-Raman and PL spectra from hot-pressed $\ce{\alpha}$-\ce{MoO_{3}} flakes on Si substrates.

\subsection{Stoichiometric characterizations}
We used Quanta 3D FIB-SEM to capture SEM images of pre- and post-processed $\ce{\alpha}$-\ce{MoO_{3}} flakes. A multipurpose JEOL-2800 transmission electron microscope with a resolution of 0.1~nm, operated at 80 to 200 kV, performed energy dispersive spectroscopy (EDS). TEM samples are prepared by immersing and ultrasonicating Si substrate capped with hotpressed $\ce{\alpha}$-\ce{MoO_{3}} flakes in ethanol for 30 minutes. This isolated the flakes in ethanol yielded an MoO$_3$/ethanol dispersion, which was drop-cast onto a carbon-coated Cu grid by a micropipette. X-ray photoelectron spectroscopy (XPS) was performed by using AXIS Supra by Kratos analytical instrument with a dual anode A1K $\alpha$ (1487.6/eV) monochromatic X-ray source, and high spatial resolution of 0.1 $\mu$m to characterize the chemical composition and stoichiometry of hot-pressed $\ce{\alpha}$-\ce{MoO_{3}}. The binding energy calibration was performed carefully by using C1s peak (284.8 eV) as a reference value.

\subsection{DFT calculations}
DFT calculations are performed using the Vienna ab initio Simulation Package (VASP)\cite{kresse1993ab} with projector augmented wave (PAW) pseudopotentials Mo (\ce{4s^{2}} \ce{4p^{6}} \ce{5s^{1}} \ce{4d^{5}}) and O (\ce{2s^{2}} \ce{2p^{4}})\cite {blochl1994projector}. We employed the vdW-DF approach, which accounts for the intriguing dispersion interaction between the structure layers. Optimizing the unit cell with various exchange functionals paired with vdW-DF. For high precision, we converged with an energy cutoff for the plane-wave basis set of 700~eV for a $3\times3\times1$ supercell integrated with a $4\times4\times2$ $\Gamma$-centered k-point grid for Brillouin zone exploration\cite{monkhorst1976special}. All structures were fully optimized until the residual forces on the ions were less than 0.01~eV for stoichiometric cells and 0.05~eV for defect cells.

\section{Supporting Information}
Details of PiFM characterization and dispersion calculations; details of FDTD simulations; details of structural and chemical stoichiometric characterization; details of Raman and XPS spectral analysis; details of DFT calculations for various pristine and hot-pressed $\ce{\alpha}$-\ce{MoO_{3}} samples. 

\section{Author Contributions}
M.R.S., N.H., and M.A.S. conceived the idea and designed the experiments. N.H. and M.A.S. prepared the samples. M.S. performed PiFM measurements. N.H. performed Raman, XPS, SEM, EDS, and GIXRD measurements. W.H. performed DFT calculations. M.A.S. performed FDTD and analytical calculations. M.A.S., N.H., and J.J.B. performed data analysis; M.A.S., N.H., and W.H. prepared figures and Supplementary Information. M.A.S., N.H., and M.R.S. drafted the manuscript, and all the co-authors contributed to its final version. M.R.S. supervised the project. 

\section{Conflict of Interest}

No conflict of interest has been identified.

\section{Data Availability}

All the data reported in this manuscript is available at (URL).

\begin{acknowledgement}
M.R.S. and W.H. acknowledge support from DARPA (grant no. D22AP00153). M.R.S. and M.S. acknowledge support from the NSF (grant no. ECCS-2339271). This research was partially supported by the National Science Foundation Materials Research Science and Engineering Center program through the UC Irvine Center for Complex and Active Materials (DMR-2011967). Micro-Raman measurements were performed in UCI Laser Spectroscopy Laboratories. M.A.S. acknowledges support from Eddleman Quantum Institute at the University of California, Irvine.

\end{acknowledgement}

\setcounter{figure}{0}    

\newpage

\section{Supporting Information}

\subsection{Nanoimaging PhPs in hot-pressed $\ce{\alpha}$-\ce{MoO_{3}} using PiFM}
\begin{figure}[ht!] 
\centering
\includegraphics[width=0.92\textwidth]{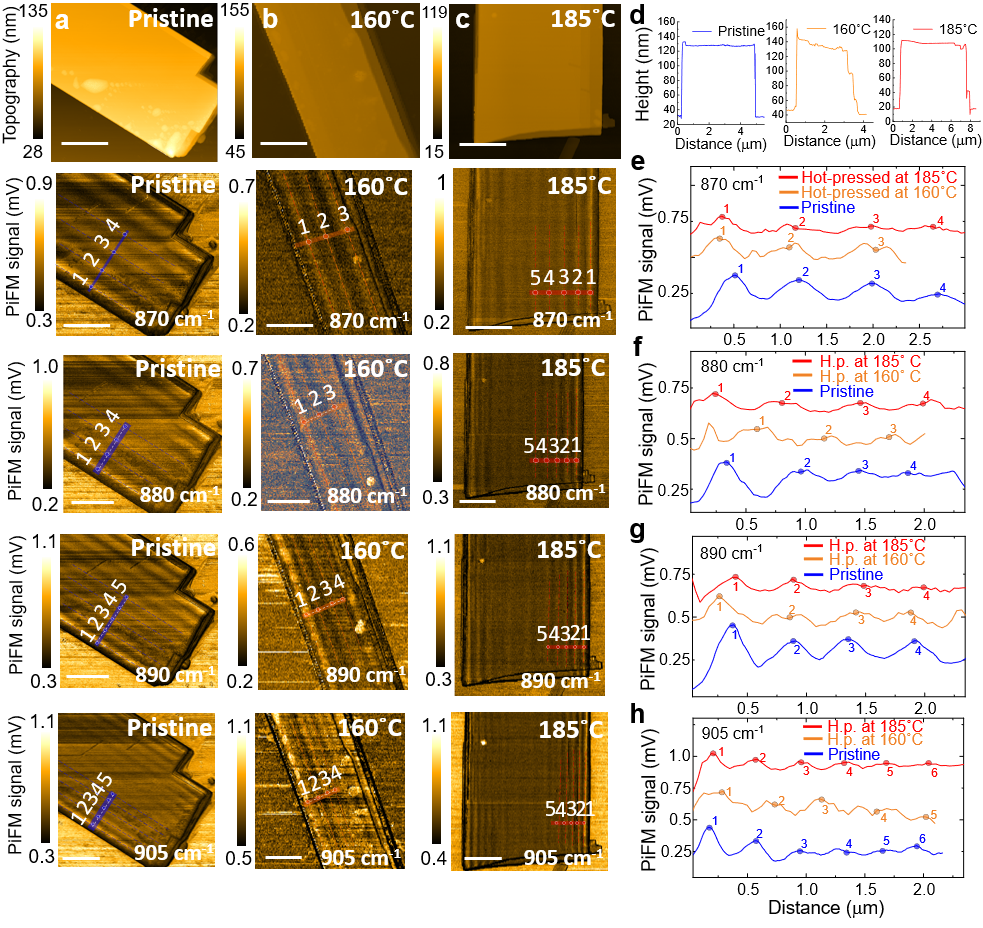}
\caption{Nanoimaging PhPs in pristine and hot-pressed $\ce{\alpha}$-\ce{MoO_{3}} on silicon substrate. Height profiles shown for (a) pristine ($t_{pr}$ = 107 nm), (b) 160$^{\circ}$ C  ($t_{h.p. 160^{\circ} C}$ = 102 nm) and (c) 185$^{\circ}$ C ($t_{h.p. 185^{\circ}  C}$ = 102 nm)  hot-pressed $\alpha$-\ce{MoO_{3}} flakes. The scales represent 1 $\mu$m. Corresponding PiFM images are shown for frequencies 870, 880, 890 and 905 $cm^{-1}$ along the columns under (a) for pristine, (b) for 160$^{\circ}$ C and (c) for 185$^{\circ}$ C h.p. $\alpha$-\ce{MoO_{3}} flakes. (d) 2D topography (height) profiles of the flakes (e-h) The extraced PhPs lineprofiles are overlapped at similar frequencies for the pristine and h.p. $\alpha$-\ce{MoO_{3}} flakes at (e) 870 $\ce{cm^{-1}}$, (f) 880 $\ce{cm^{-1}}$, (g) 890 $\ce{cm^{-1}}$, and (h) 905 $\ce{cm^{-1}}$. (e-h) shows that hot-pressing elongates the PhP wavelength ($\lambda_{PhP}$) at each of these frequencies shown suggesting the $\alpha$-\ce{MoO_{3}} L-RB PhP dispersion to be tuned towards lower momenta regime.}
\label{fig2}
\end{figure}

\subsection{Extraction of PhP propagation characteristics in $\alpha$-\ce{MoO_{3}}}
We performed the lifetime calculation of $\alpha$-\ce{MoO_{3}} PhPs according to $\tau_x$=\ce{$L_{X}$}/$v_g$. Here, the group velocity is calculated from the experimentally recorded dispersion curves (shown in Fig. 2c-f). 
Group velocity is defined by $v_g$ = \textit{d}$\omega$ / \textit{d}$\ce{k}_{x}$. In order to calculate the group velocity of the $\alpha$-MoO3 PhPs, we use the first-order derivative of the dispersion curves in Fig. 3b (main text) which we get from the experimental PiFM measurements. We take the derivative along the [100] crystal axis directions since our frequency range of interest lies in the L-RB. We numerically fit the experimental dispersion curve data points by a general potential function. This function is taken to be of y= $\ce{ax}^{b}$. Followed by the fitting, to get $v_g$, we calculate a numerical derivative of the resulting curves.

Moreover, for extracting the propagation length (\ce{$L_{X}$}), we fitted the near-field line profiles of the PhPs. These near-field line profiles represent the real part of the z-component of the electric fields along the crystal direction of [100]. We fitted the captured PiFM signal to an exponentially decaying sinusoidal signal along with a dissipation factor. As the PhPs start to propagate across the distance x, the field starts to decay exponentially. The model equation upon which we performed our fitting is mentioned below\cite{ma2018plane}: 

\begin{equation}
y =  y_{0} + \ce{A} e^{(-x/t_{0})}  \sin (\pi \frac{\ce{x}-\ce{x_{c}}}{\ce{w}})
\end{equation}

where A$>$0, W$>$0 and $t_0$ $>$0. For the fitting procedure, we used the Levenberg-Marquardt iteration algorithm. After fitting is performed, we can calculate from the fitted parameter $t_0$ which represents an estimate of \ce{$L_{X}$} of PhPs.




\newpage
\subsection{PiFM PhP characterization of pristine and h.p. $\alpha$-\ce{MoO_{3}} flakes at 160$^{\circ}$C and 185$^{\circ}$C}
\begin{figure}[ht!] 
\centering
\includegraphics[width=1\textwidth]{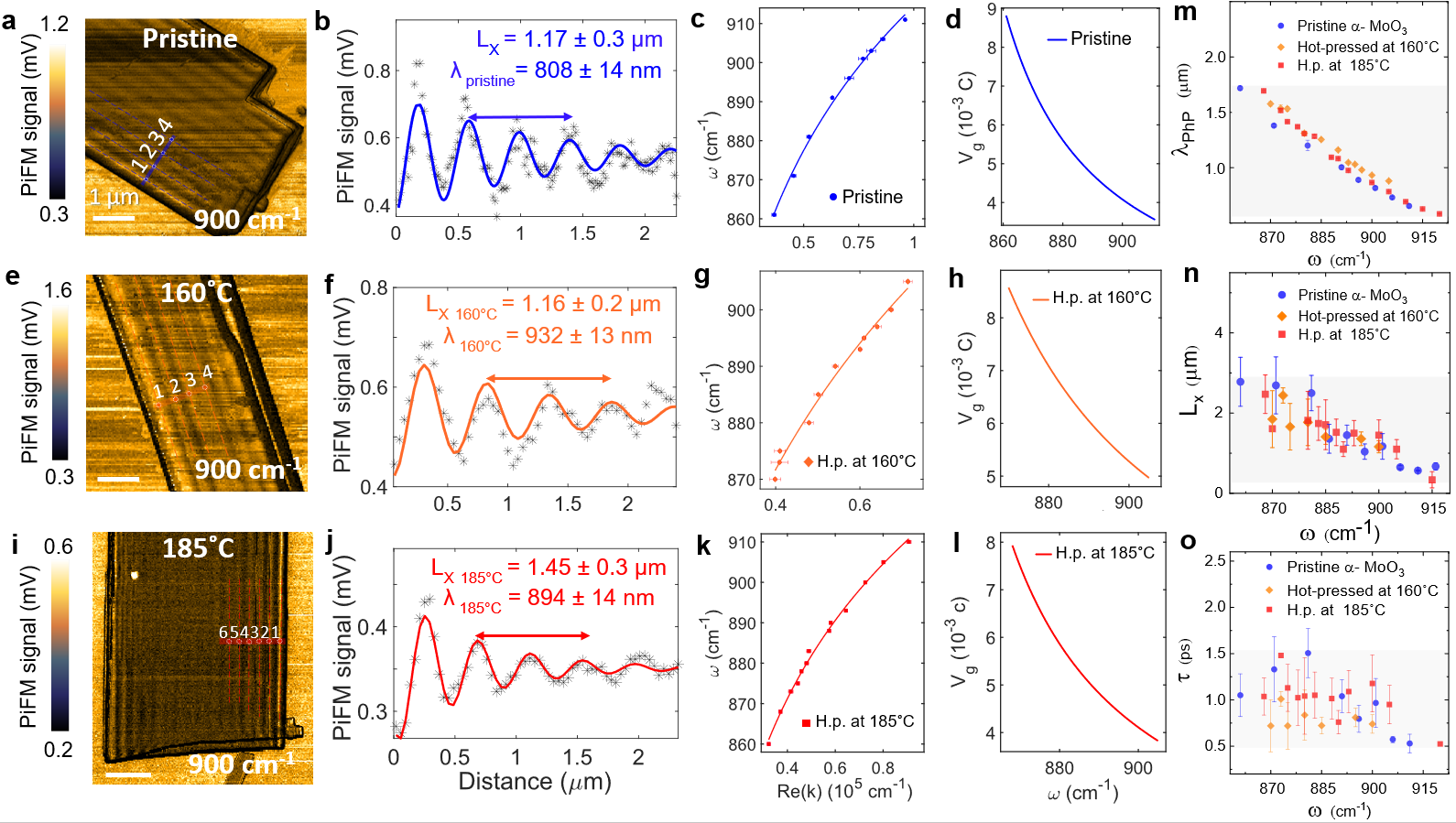}
\caption{ PhP characterization of pristine and h.p. $\alpha$-\ce{MoO_{3}} using PiFM. Extracted PhP lineprofiles are fitted to an exponentially-decaying sinusoidal signal model for (a-b) pristine and h.p. $\alpha$-\ce{MoO_{3}} flakes at (e-f) 160$^{\circ}$C and (i-j) 185$^{\circ}$C. As a representation, the fittings for 900 $\ce{cm^{-1}}$ is shown here.  The dispersion and group-velocity for each of these cases are shown for (c-d) pristine and h.p. flakes at (g-h) 160$^{\circ}$C and (k-l) 185$^{\circ}$C, respectively. The key figure of merits used for characterizing the PhP propagation dynamics are PhP wavelength ($\lambda_{PhP}$), propagation length (\ce{$L_{X}$}) and lifetime ($\tau$). The comparisons in these FOMs are shown in (m) for $\lambda_{PhP}$, (n) for \ce{$L_{X}$} and (o) for $\tau$ of pristine and h.p. $\alpha$-\ce{MoO_{3}} flakes. The error bars of lifetimes were calculated based on the errors of propagation lengths originating from the standard deviation of fitting. Measurements are performed within the lower Reststrahlen band (L-RB) of $\alpha$-\ce{MoO_{3}}, ranging from 865 to 915 $\ce{cm^{-1}}$.}
\label{fig2}
\end{figure}

\newpage
\subsection{PiFM PhP characterization of h.p. $\alpha$-\ce{MoO_{3}} flake at 200$^{\circ}$C}
\begin{figure} 
\centering
\includegraphics[width=1\textwidth]{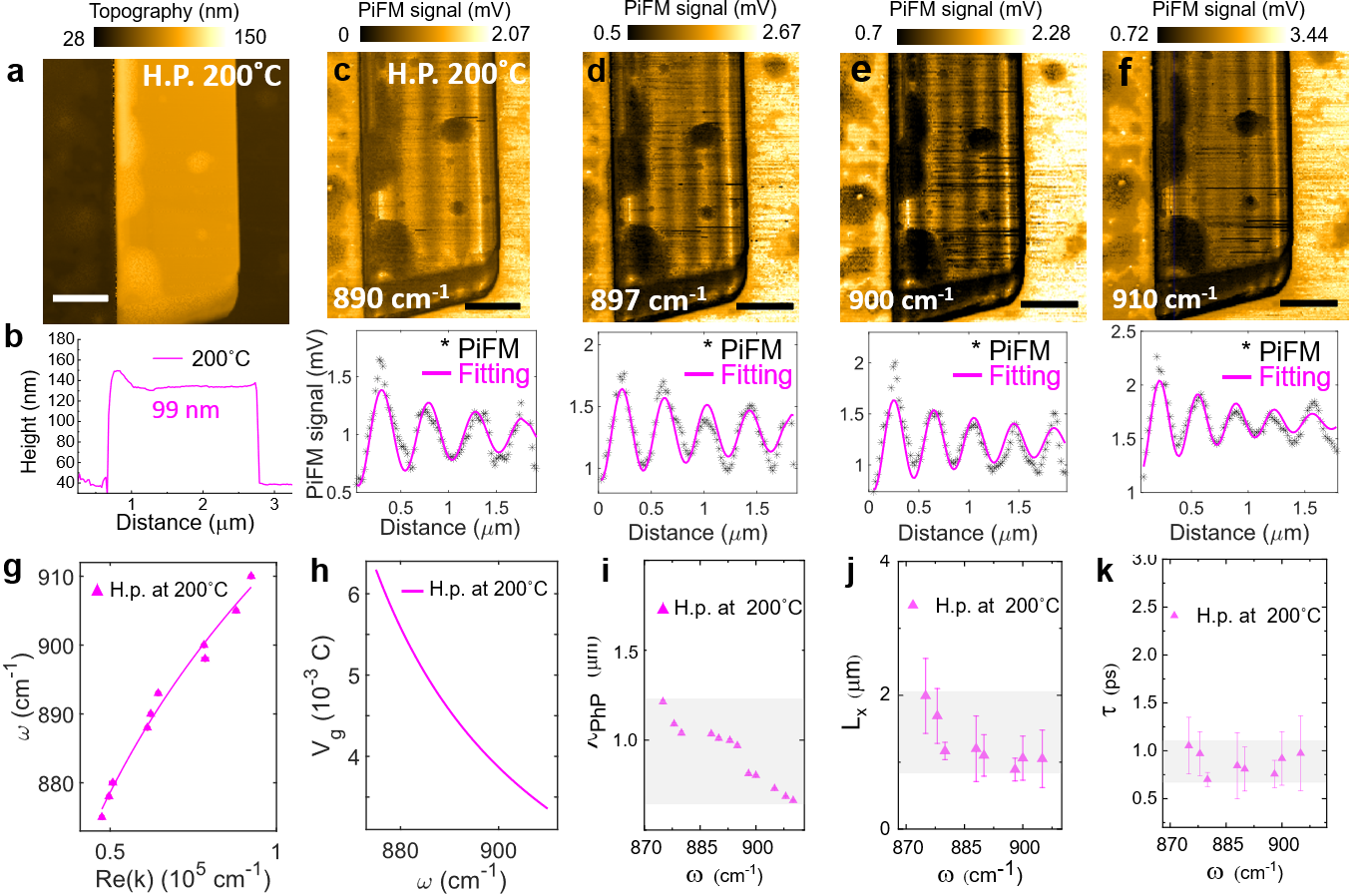}
\caption{Nanoimaging PhPs in 200$^{\circ}$C hot-pressed $\alpha$-\ce{MoO_{3}} using PiFM. (a-b) Height profile shown for the measured h.p. flake with thickness, $t_{h.p. 185^{\circ} C}$ = 99 nm. Recorded PiFM images are shown for representative frequencies of (c) 890, (d) 897, (e) 900 and (f) 910 $\ce{cm^{-1}}$. Corresponding PhP lineprofiles are fitted to an exponentially-decaying sinusoidal function as shown in (c-f). The dispersion and group-velocity for 200$^{\circ}$ C hot-pressed $\alpha$-\ce{MoO_{3}} are shown in (g) and (h), respectively. The FOMs for PhP propagation are shown over frequencies in (i-k). The error bars of lifetimes are calculated based on the errors of propagation lengths originating from the standard deviation of fitting. Measurements are performed within the lower Reststrahlen band (L-RB) of $\alpha$-\ce{MoO_{3}}, ranging from 865 to 915 $\ce{cm^{-1}}$.}
\label{fig2}
\end{figure}


\newpage
\subsection{Figure of merits of thermomechanically processed $\alpha$-\ce{MoO_{3}} flakes}
\begin{figure} 
\centering
\includegraphics[width=1\textwidth]{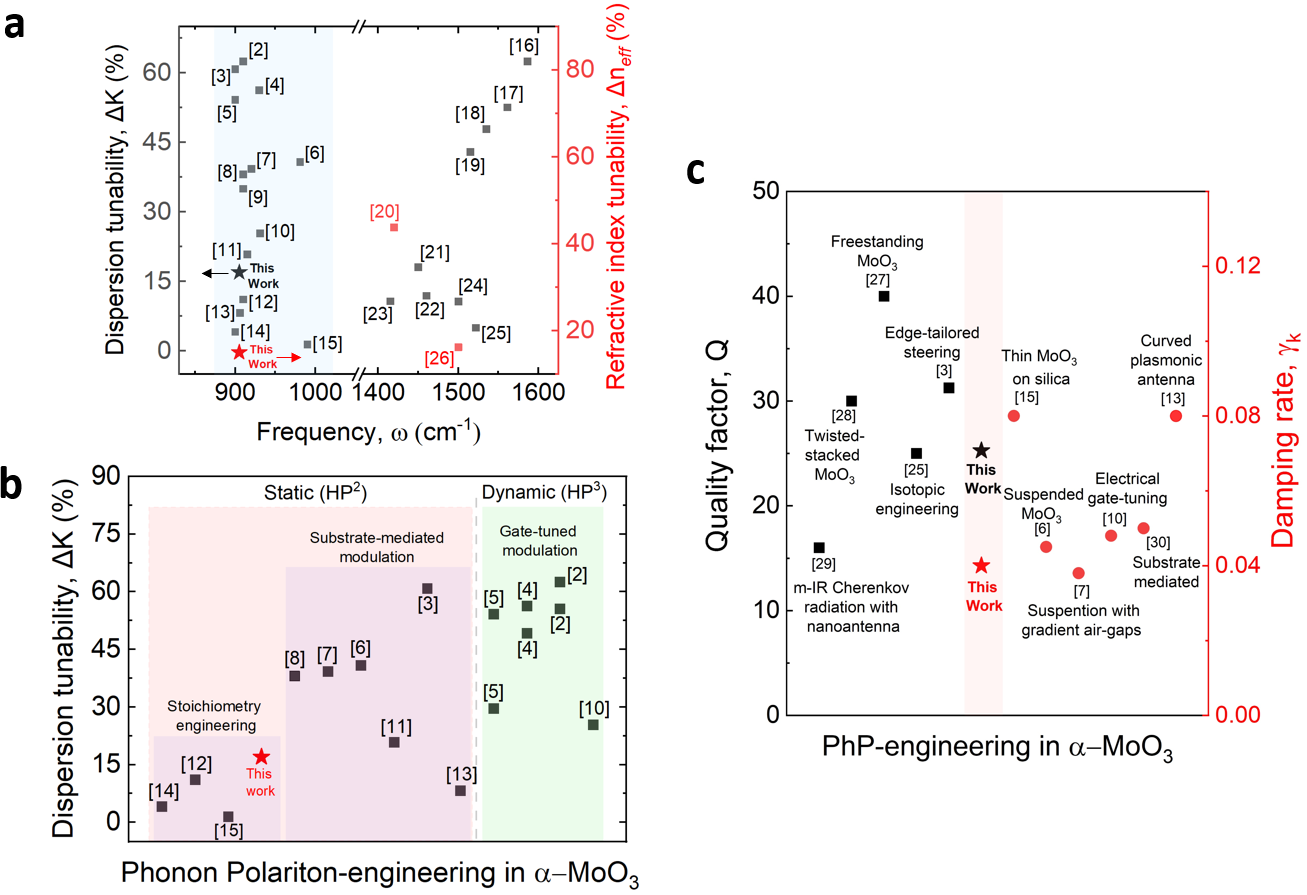}
\caption{Figure of merit analysis of thermomechanically-processed $\alpha$-\ce{MoO_{3}}. (a) Dispersion tunability of PhPs extracted from reported works that are related to IR-nanoimaging of PhPs across different van der Waals (vdW) materials. (b-c) Comparison of this work with respect to quality factor (Q) and PhP damping rate ($\gamma_{k}$) with respect to the reported PhP engineering mechanisms \cite{hu2022doping,dai2020edge,alvarez2022active, zeng2022tailoring, shen2022hyperbolic, zheng2022tunable, schwartz2021substrate, ruta2022surface, zhou2023gate, yang2022high, zhao2022ultralow, zheng2022controlling, wu2020chemical, zheng2018highly, fali2019refractive, kim2017effect, dai2015graphene, he2021guided, chaudhary2019engineering, folland2018reconfigurable, dai2018hyperbolic, wang2020probing, virmani2021amplitude, ni2021long, chaudhary2019polariton, yang2022high, teng2024steering, guo2023mid, zhang2021hybridized} in $\alpha$-\ce{MoO_{3}}.} 
\label{fig2}
\end{figure}

The figure of merit (FOM) of PhPs in $\alpha$-\ce{MoO_{3}} are calculated along the [100] crystal direction. For analysis of PhP FOMs, we calculate the the Q-factor (Q) using the supplementary equation 2\cite{ni2021long,dai2020edge,yang2022high}. 

\begin{equation}
Q =   \frac{\ce{Re($\ce{k}_{x}$)}}{\ce{Im($\ce{k}_{x}$)}}
\end{equation}

We calculate the \ce{Re($\ce{k}_{x}$)} and \ce{Im($\ce{k}_{x}$)} by fitting the PiFM linescans as mentioned in the previous section. Here, the \ce{Im($\ce{k}_{x}$)} is related to the propagation length (\ce{$L_{X}$}) by the supplementary equation 3\cite{ni2021long,dai2020edge,yang2022high}.

\begin{equation}
\ce{Im($\ce{k}_{x}$) =   \frac{\ce{$1$}}{2 \ce{$L_{X}$}}}
\end{equation}

\subsection{Dielectric permittivity modeling of pristine and hot-pressed $\alpha$-\ce{MoO_{3}} in FDTD} 
\begin{figure} [ht!]
\centering
\includegraphics[width=1\textwidth]{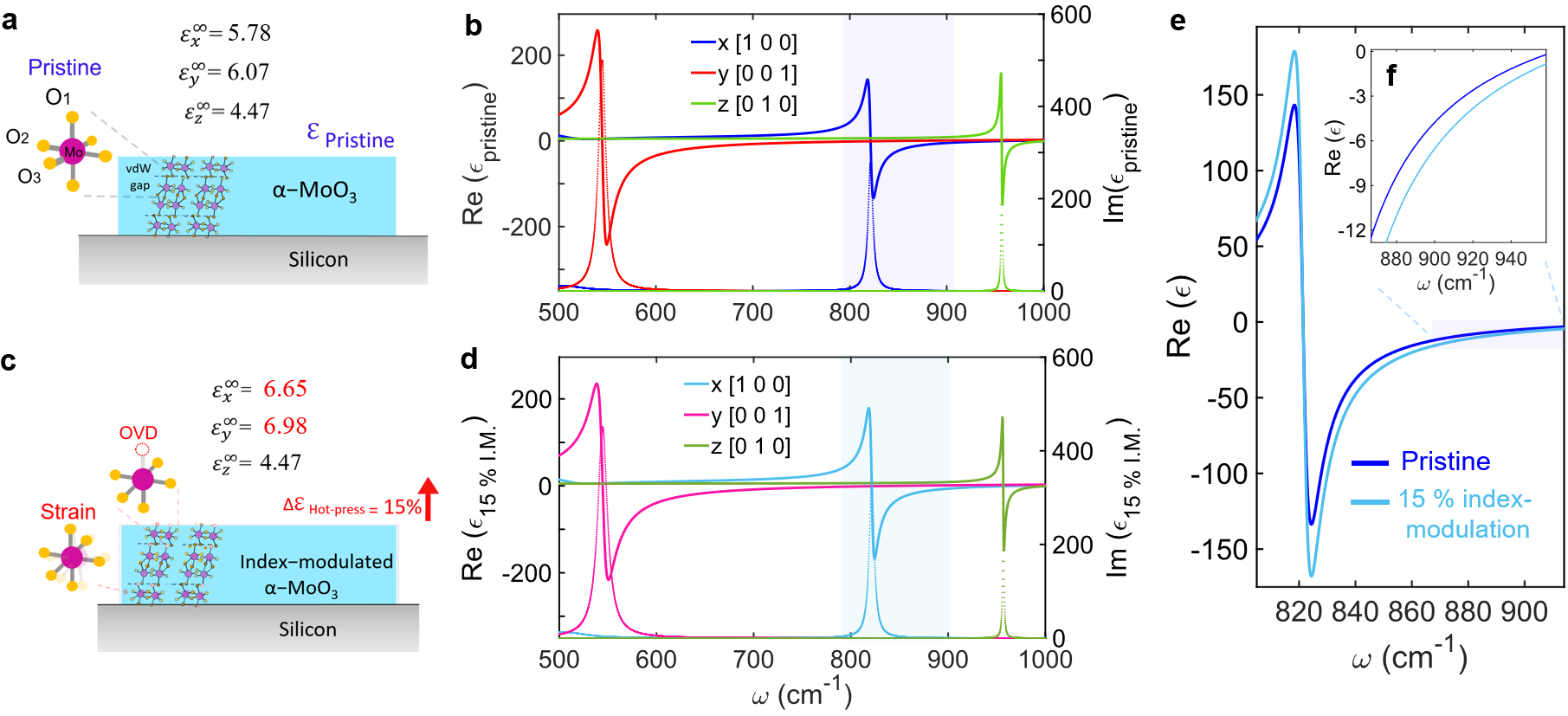}
\caption{Dielectric permittivity modeling of pristine and hot-pressed $\alpha$-\ce{MoO_{3}} in FDTD. (a) Schematic illustration of a pristine $\alpha$-\ce{MoO_{3}}. (b) The dielectric permittivities for this pristine $\alpha$-\ce{MoO_{3}} model are plotted along the x [100], y [001] and z [010]. The crystallographic axes: [001], [100] and [010], correspond to phonons in each of the Reststrahlen bands ($\ce{RB_{1} (red)}$ , $\ce{RB_{2} (blue)}$  and $\ce{RB_{3} (green)}$, respectively. (c-d) Hot-pressed $\alpha$-\ce{MoO_{3}} is modeled by introducing the effects of strain and OVD through tuning the static dielectric constants along the $x$ and $y$ crystal-direction. Schematic illustration of index-modulated;  h.p. $\alpha$-\ce{MoO_{3}} model is shown in (c). The dielectric permittivities for this index-modulated $\alpha$-\ce{MoO_{3}} model are plotted in (d). In (b-d), the region of interest over the L-RB is shown with shaded areas. (e) Changes in the L-RB dispersion relationships (shown in the shaded regions in (b-d)) are overlapped for analyzing the spectral tuning of the L-RB.  The inset shows a zoomed-in look into the shaded region of interest from 865-915 $\ce{cm^{-1}}$ between the pristine and index-modulated $\alpha$-\ce{MoO_{3}} model.}
\label{fig3}
\end{figure}

Since $\alpha$-\ce{MoO_{3}} is an anisotropic polar dielectric crystal, we model the permittivity by the Lorentz model for the case of coupled oscillators (also known as the TO-LO model) \cite{alvarez2020infrared,alvarez2019analytical,ma2018plane,folland2018reconfigurable}. We use three oscillators for approximating $\epsilon_{x}$ and one oscillator each for the cases of $\epsilon_{y}$ and $\epsilon_{z}$, respectively.

\begin{equation}
\epsilon_{x}(\omega)=\epsilon_{x}^\infty 
\begin{pmatrix}
\frac{(\omega_{x1}^{LO})^2-\omega^2-i\gamma_{x1}\omega}{(\omega_{x1}^{TO})^2-\omega^2-i\gamma_{x1}\omega}
\end{pmatrix}
\begin{pmatrix}
\frac{(\omega_{x2}^{LO})^2-\omega^2-i\gamma_{x2}\omega}{(\omega_{x2}^{TO})^2-\omega^2-i\gamma_{x2}\omega}
\end{pmatrix}
\begin{pmatrix}
\frac{(\omega_{x3}^{LO})^2-\omega^2-i\gamma_{x3}\omega}{(\omega_{x32}^{TO})^2-\omega^2-i\gamma_{x3}\omega}
\end{pmatrix}
\end{equation}

\begin{equation}
\epsilon_{y}(\omega)=\epsilon_{y}^\infty 
\begin{pmatrix}
\frac{(\omega_{y1}^{LO})^2-\omega^2-i\gamma_{y1}\omega}{(\omega_{y1}^{TO})^2-\omega^2-i\gamma_{y1}\omega}
\end{pmatrix}
\end{equation}

\begin{equation}
\epsilon_{z}(\omega)=\epsilon_{z}^\infty 
\begin{pmatrix}
\frac{(\omega_{y1}^{LO})^2-\omega^2-i\gamma_{y1}\omega}{(\omega_{y1}^{TO})^2-\omega^2-i\gamma_{y1}\omega}
\end{pmatrix}
\end{equation}

In this Lorentz model, the three principal axes of the $\alpha$-\ce{MoO_{3}} can be considered in the [100], [001] and [010] directions. Here, the $\epsilon_{x}$ ($\omega$), $\epsilon_{y}$ ($\omega$), and $\epsilon_{z}$ ($\omega$) represent the three principal components of the permittivity tensor, respectively. To generalize this representation, the permittivity tensors can be then denoted by $\epsilon_{i}$ ($\omega$) where the $i = x,y,z$. Here, the static dielectric constant is represented by $\epsilon_{i}^\infty$. The LO and TO phonon frequencies along the three directions is represented by $\omega_{ij}$(LO) and $\omega_{ij}$(TO) along the $i$-th direction with $i= x,y,z$ respectively. Moreover, $\gamma_{ij}(LO)$ represent the damping factor of the Lorentzian line shape derived from the phonon scattering rate. The subscript $j$ here represents the different phonon pairs along the same axis. The parameters used in modeling the dielectric permittivity model of the $\alpha$-\ce{MoO_{3}} system have been utilized by model fitting the optical response of the material measured from polarized far-field IR spectroscopy\cite{alvarez2020infrared}. The static dielectric permittivity tensor components are taken as $\epsilon_{x}^\infty$ = 5.78,  $\epsilon_{y}^\infty$ = 6.07, and $\epsilon_{z}^\infty$ = 4.47.

We run density functional theory (DFT) calculations for the hypothesis of thermomechanically index-modulated $\alpha$-\ce{MoO_{3}}. The DFT calculations reflect an increase in the static dielectric constants for a hot-pressed flake. We use FDTD to simulate the influence of lattice strain- and OVD-induced index-modulation over PhPs. The thermomechanical-index modulation is mimicked by tuning the dielectric permittivity. As suggested from our DFT calculations, we increase the static dielectric permittivity tensor components, $\epsilon_{x}^\infty$  and $\epsilon_{y}^\infty$, along the [100] and [001] directions, respectively. This modifies the L-RB for the hot-pressed $\alpha$-\ce{MoO_{3}}. We show this modified L-RB calculation for the 160$^{\circ}$C hot-pressed case. For this case, the static dielectric constants are used as, $\epsilon_{x}^{h.p.}$= 1.15 $\times$ $\epsilon_x^\infty$= 6.65 and $\epsilon_{y}^{h.p.}$= 1.15 $\times$ $\epsilon_{y}^\infty$ = 6.98. It is worth noting that the tensor component along the [001] direction shows insignificant changes from the DFT calculations, as the van der Waals gap limits interlayer interaction. As a result, the static dielectric constant $\epsilon_{z}^\infty$ remains unchanged. Further, we chose to neglect the LO/TO frequency shifts in our modeling, as such calculations for large defect structures are computationally expensive and may not be physically relevant given the periodic nature of the defects.

\subsection{PiFM and FDTD for a 107 nm pristine $\alpha$-\ce{MoO_{3}} flakes on a silicon substrate.}
\begin{figure}[ht!]
\centering
\includegraphics[width=1.\textwidth]{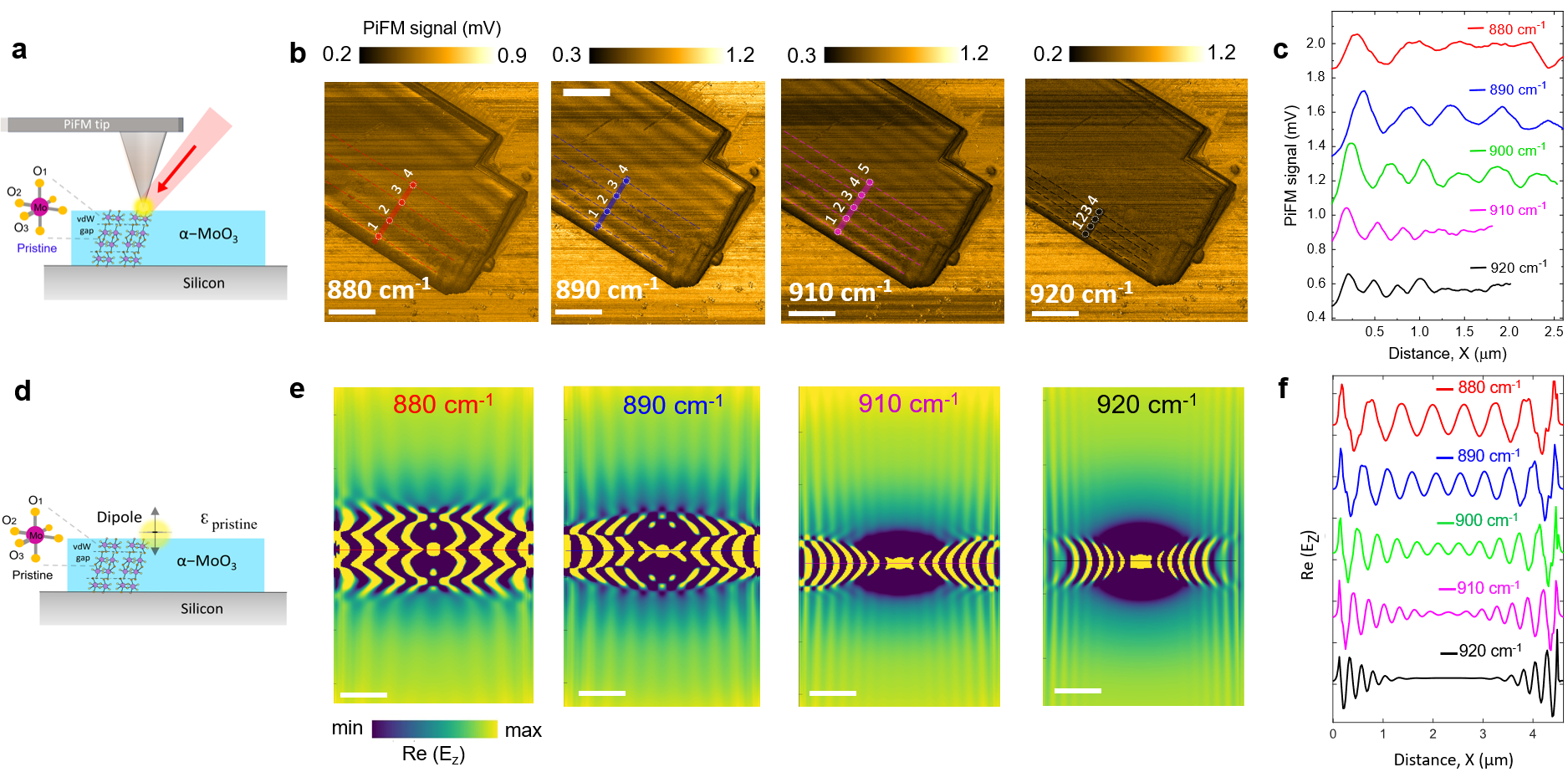}
\caption{ PiFM and FDTD for a 107 nm pristine $\alpha$-\ce{MoO_{3}} flakes on a silicon substrate. (a) Schematic illustration of PiFM (b) PiFM images and (c) extracted PhP linescans with for excitations with frequencies from 860 – 920 $\ce{cm^{-1}}$ within the L-RB of $\alpha$-\ce{MoO_{3}}. The scale bar represents 2 $\mu$m. (d) Schematic illustration of dipole-launched PhPs for FDTD (d) Numerically simulated out-of-plane component of electric-field (Re($Ez$)) for frequencies from 860 – 920 $\ce{cm^{-1}}$ and (e) corresponding linescan profiles for the PhPs}
\label{fig1}
\end{figure}


\newpage
\subsection{Analytical model for tunable dispersion of thermomechanically-engineered $\alpha$-\ce{MoO_{3}}}
\begin{figure}[ht!]
\centering
\includegraphics[width=1\textwidth]{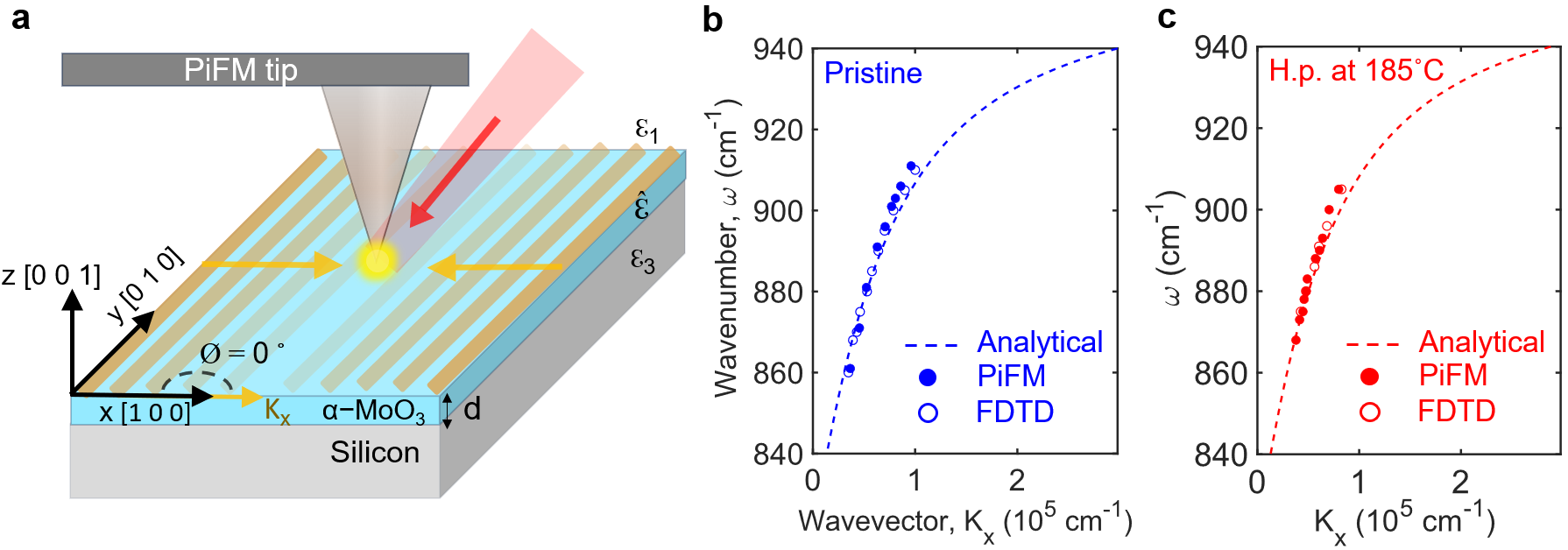}
\caption{ Analytical modeling for dispersion of PhPs in thermomechanically-engineered $\alpha$-\ce{MoO_{3}} slabs (a) Schematics of the $\alpha$-\ce{MoO_{3}} biaxial slab on a silicon substrate. Crystal axes [1 0 0] (directed along $x$) and [0 1 0] (directed along $y$) belong to a plane parallel to the faces of the $\alpha$-\ce{MoO_{3}} slab. The crystal axis [0 0 1] coincides with the $z$ axis. To generalize the hyperbolic L-RB mode propagation, the wave vector K propagates at an angle $\phi$ with respect to the $x$ (along [1 0 0]) axis. Here, PiFM tip-launched PhP modes originate from the dipole-like excitation source that is originated at the tip due to the mid-IR L-RB excitation frequencies. The PhP modes propagate towards the edge of the flake and gets reflected from the edge. This creates the PhP fringe pattern that propagates the PhP modes along the $x$ [1 0 0] direction. (b-c) Analytically calculated (dashed lines) dispersion relationships are plotted along with PiFM (solid circles) and full-wave FDTD calculations (empty circles) for (b) pristine (blue) and (c) 185$^{\circ}$ C hot-pressed $\alpha$-\ce{MoO_{3}} flakes (red), respectively.}
\label{fig1}
\end{figure} 

we consider a $\alpha$-\ce{MoO_{3}} slab as a biaxial medium with a general dielectric permittivity tensor $\hat{\epsilon}$. In general the $\hat{\epsilon}$ takes a form of\cite{alvarez2019analytical} 

\begin{equation}
\hat{\epsilon} (\omega) =
\begin{pmatrix}

\epsilon_x (\omega) & 0 & 0\\
0 & \epsilon_y (\omega) & 0 \\
0 & 0 & \epsilon_z (\omega)
\end{pmatrix}
\end{equation}

In Fig. S7 (a), the schematic shows the PhP modes launched by the PiFM tip as it is excited with mid-IR L-RB excitation frequencies. Here, $\ce{k}_{0}$ = $\omega$/\textit{c} is the free-space wave vector. Within the L-RB band, $\ce{k}_{x}$ makes an angle of $\phi$ = 0$^{\circ}$ with PhP modes that is propagating along the direction x [1 0 0] axis. Here, $\ce{k}_{x,y}$ is the in-plane momentum along the crystal axes $x$ (specifically along [1 0 0]) and $y$ (specifically along [0 1 0]) direction, respectively.   

For the fundamental modes within the L-RB, the PhP fringes parallel to the [001] direction satisfy the Fabry-P$\tilde{e}$rot quantization condition \cite{alvarez2019analytical, ma2018plane, shen2022hyperbolic}. In general, we calculate the wavevector that satisfies the Fabry-P$\tilde{e}$rot condition based on an analytical solution as \cite{alvarez2019analytical},

\begin{equation}
q = \frac{\rho}{d K_0}[\arctan\frac{\epsilon_1 \rho}{\epsilon_z} + \arctan\frac{ \epsilon_3\rho}{\epsilon_z}+\pi l] 
\end{equation}

Here, $l$ = 0 for the case of the fundamental mode propagation. Moreover, $q_{x,y}$ are the in-plane components of the normalized wave vector in the form of $q_{x,y}$ = $\ce{k}_{x,y}$/$\ce{k}_{0}$. $\rho$ can be defined as 

\begin{equation}
\rho = i \sqrt{\frac{\epsilon_z q^2}{\epsilon_x {q_x}^2 + \epsilon_y {q_y}^2}} = i \sqrt{\frac{\epsilon_z}{\epsilon_x \cos^2\phi + \epsilon_y \sin^2 \phi}}
\end{equation}

For the range of L-RB excitation frequencies with $\phi$ = 0$^{\circ}$, equation (4) takes a form of   

\begin{equation}
\rho = i \sqrt{\frac{\epsilon_z}{\epsilon_x}} 
\end{equation}



\newpage
\subsection{SEM and Raman characterization of pristine and hot-pressed $\alpha$-\ce{MoO_{3}} } 
\begin{figure} [ht!]
\centering
\includegraphics[width=0.9\textwidth]{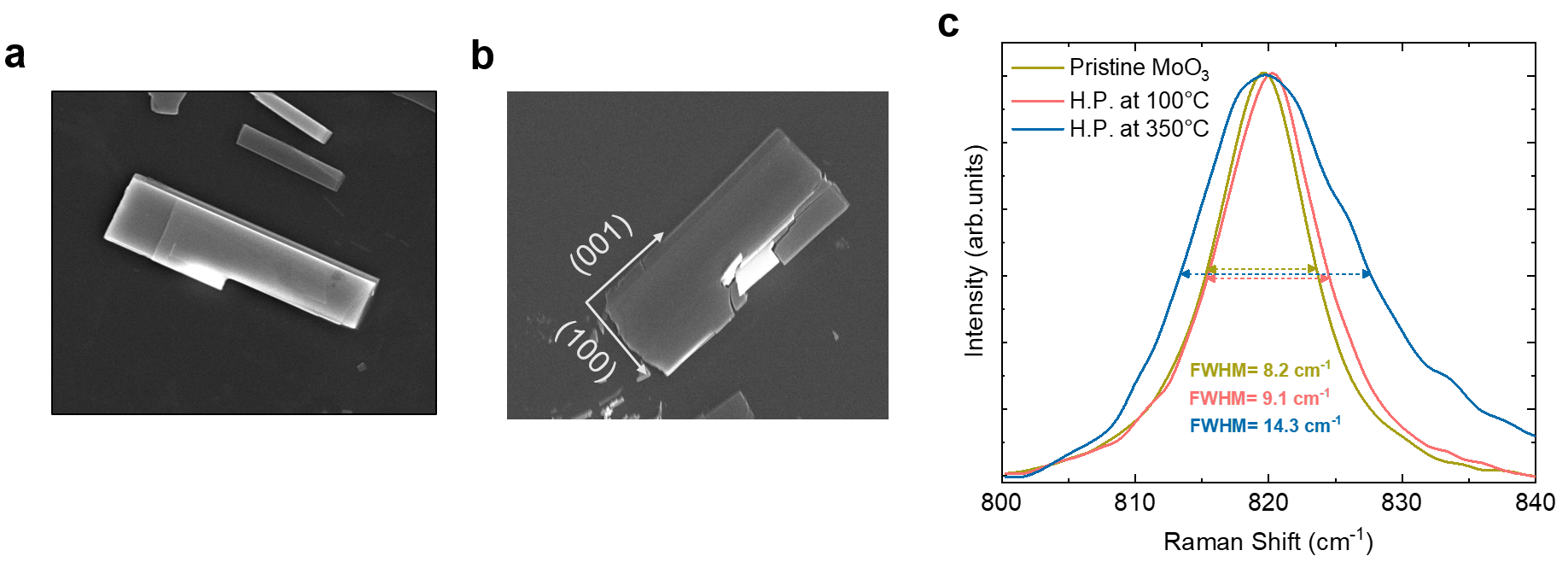}
\caption{ SEM and Raman characterization of pristine and hot-pressed $\alpha$-\ce{MoO_{3}}. (a-b) SEM image of an exfoliated (a) pristine $\alpha$-\ce{MoO_{3}} flake and (b) hot-pressed flake with its crystal orientation. The scale bar is 10 $\mu$m. (c) Comparison of the FWHMs of $\ce{A}_{\ce{g}}$/$\ce{B}_{\ce{1g}}$ phonon mode of pristine and h.p. $\alpha$-\ce{MoO_{3}} at 150$^{\circ}$C and 350$^{\circ}$C. The increase in FWHM with increase in h.p. temperature indicates the introduction of vacancy defect states in $\alpha$-\ce{MoO_{3}}}
\label{fig3}
\end{figure}


\subsection{X-ray photolectron spectroscopy (XPS) suvey} 
\begin{figure} [ht!]
\centering
\includegraphics[width=0.7\textwidth]{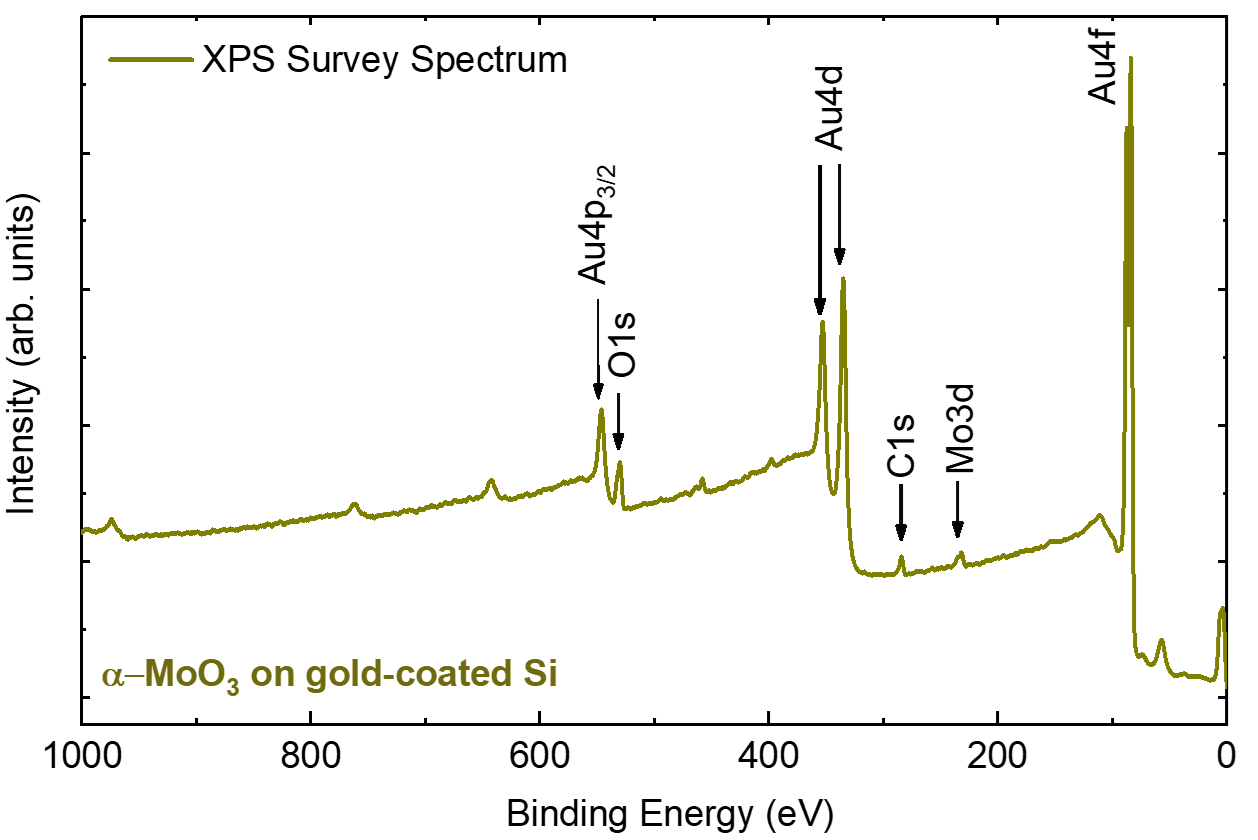}
\caption{X-ray photoelectron spectroscopy (XPS) survey spectrum of h.p. $\alpha$-\ce{MoO_{3}} at 350$^{\circ}$C on gold-coated substrate.}
\label{fig3}
\end{figure}


\newpage
\subsection{Energy dispersive spectroscopy (EDS) measurements} 
\begin{figure} [ht!]
\centering
\includegraphics[width=0.75\textwidth]{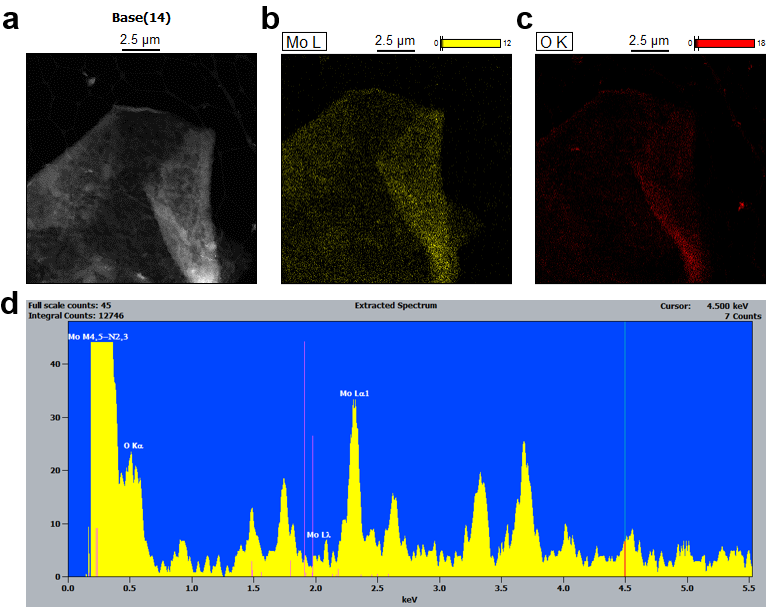}
\caption{EDS measurement of $\alpha$-\ce{MoO_{3}} hot-pressed at 350$^{\circ}$C. (a) TEM image of the flake. (b-c) elemental mapping of Mo and O2 collected from an area highlighted in (a). (d) Elemental intensity profile of mapping.}
\label{fig3}
\end{figure}

\subsection{EDS-mapped reduced oxidation state in hot-pressed $\alpha$-\ce{MoO_{3}}} 
\begin{figure} [ht!]
\centering
\includegraphics[width=1\textwidth]{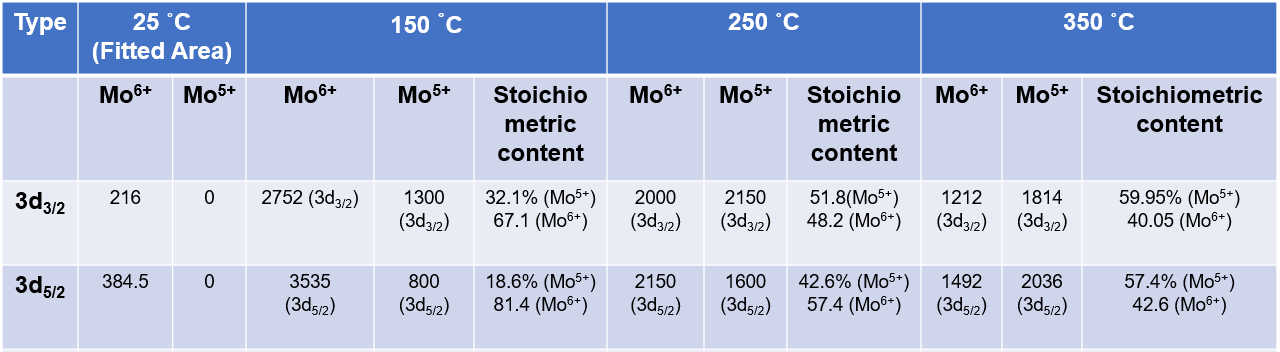}
\caption{EDS-mapped reduced oxidation state in hot-pressed $\alpha$-\ce{MoO_{3}} is extracted by spectral peak-fitting of the EDS elemental intensity map.}
\label{fig3}
\end{figure}

\newpage
\subsection{DFT calculations}
DFT calculations were performed using the Vienna Ab initio Simulation Package (VASP) \cite{kresse1996efficient,kresse1993ab} with projector augmented wave (PAW) pseudopotentials for Mo (4s2 4p6 5s1 4d5) and O (2s2 2p4). The vdW-DF approach was employed to account for dispersion interactions between the layers. A Hubbard U term of 5 eV was applied to Mo to better account for the localized d-electrons. Convergence was achieved with an energy cutoff of 700 eV for the plane-wave basis set with a 9×9×3 gamma-centered k-mesh\cite{monkhorst1976special,langreth2005van,tong2021first} for stoichiometric cell calculations and 4x4x3 for the 3x3x1 super-cell structures. All structures were optimized until the residual forces on the ions were less than 0.01 eV for stoichiometric cells and 0.05 eV for defect cells. The static dielectrics are obtained using density functional perturbation theory within VASP\cite{langreth2005van,tong2021first}.


\subsection{DFT calculations of OVD-induced index modulation in $\alpha$-\ce{MoO_{3}}}
\begin{figure} [ht!]
\centering
\includegraphics[width=0.95\textwidth]{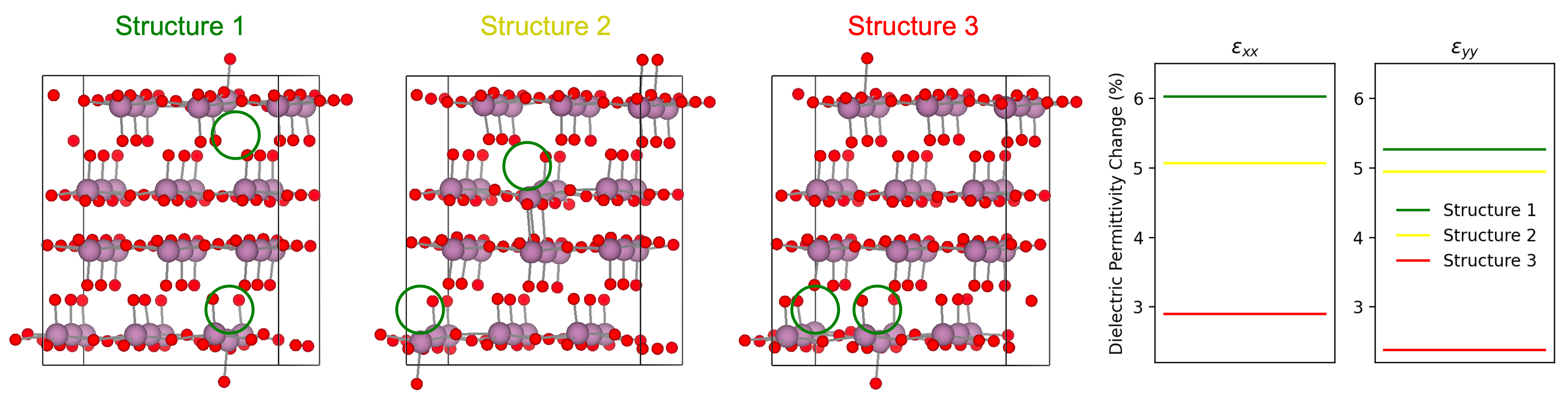}
\caption{Structural and dielectric changes in $\alpha$-\ce{MoO_{3}} with two O1 site defects placed at different relative positions near the vdW gap. Structures 1, 2, and 3 correspond to distinct randomly selected defect configurations. The dielectric permittivity change (in \%) relative to the pristine structure is shown for the $\mathrm{\epsilon_{xx}}$ and $\mathrm{\epsilon_{yy}}$ components, highlighting the impact of defect placement on the material's dielectric response. In particular, Structure 3, where vacancies are nearby and located on the same layer, exhibits much smaller modulation, while Structures 1 and 2 yield similar results. Structure 1 is used in the main text as it is less effected by vacancy proximity.}
\label{fig5}
\end{figure}


\newpage
\subsection{DFT band structure of strained $\alpha$-\ce{MoO_{3}}}
\begin{figure} [ht!]
\centering
\includegraphics[width=0.95\textwidth]{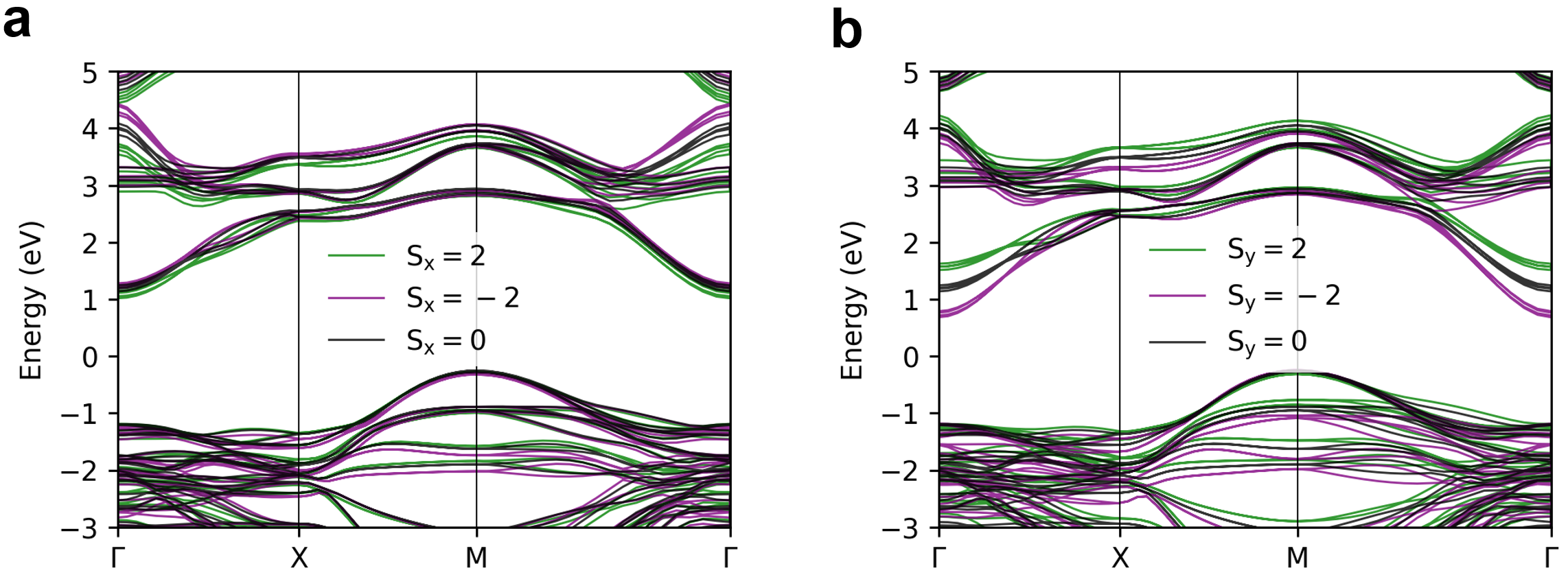}
\caption{Band structure of strained MoO$_3$ for positive and negative strains. The left panel shows results for strain along x ($S_x = 2$, $S_x = -2$, and $S_x = 0$), while the right panel shows strain along y ($S_y = 2$, $S_y = -2$, and $S_y = 0$). Notably, the bands near the $\Gamma$-point exhibit significant shifts under $S_y$ strain, altering the band gap. This shift correlates with changes in the static dielectric constant, as noted in the main text, where a smaller band gap leads to larger static dielectric.}
\label{fig6}
\end{figure}

\providecommand{\latin}[1]{#1}
\makeatletter
\providecommand{\doi}
  {\begingroup\let\do\@makeother\dospecials
  \catcode`\{=1 \catcode`\}=2 \doi@aux}
\providecommand{\doi@aux}[1]{\endgroup\texttt{#1}}
\makeatother
\providecommand*\mcitethebibliography{\thebibliography}
\csname @ifundefined\endcsname{endmcitethebibliography}
  {\let\endmcitethebibliography\endthebibliography}{}


\end{document}